\documentclass[12pt]{iopart}
\usepackage{iopams} 
\usepackage{cite} 
\usepackage{xcolor}
\usepackage{graphicx}
\usepackage{ntheorem}
\usepackage{textgreek} 
\usepackage{hyperref}
\hypersetup{
  colorlinks=true,
  urlcolor=blue,
  linkcolor=blue,
  pdftoolbar=true,
  pdfmenubar=true,
  citecolor=blue,
  filecolor=blue
}
\usepackage{soul}

\newtheorem*{theorem*}{Theorem}
\newtheorem*{definition*}{Definition}

\eqnobysec

\newcommand{\tauc}{\tau_{\mathrm{cap}}}

\begin{document}

\title[Stochastic 1D search-and-capture as a $G/M/c$ queueing model]{Stochastic 1D search-and-capture as a G/M/c queueing model}

\author{Jos\'{e} Giral-Barajas and Paul C Bressloff}

\address{Department of Mathematics, Imperial College London, London SW7 2AZ, UK}
\ead{\mailto{j.giral-barajas24@imperial.ac.uk}, \mailto{p.bressloff@imperial.ac.uk}}
\vspace{10pt}
\begin{indented}
\item[] August 2025
\end{indented}

\begin{abstract}
We study the accumulation of resources within a target due to the interplay between continual delivery, driven by 1D stochastic search processes, and sequential consumption. The assumption of sequential consumption is key because it changes the commonly used $G/M/\infty$ queue to a $G/M/c$ queue. Combining the theory of $G/M/c$ queues with the theory of first-passage times, we derive general conditions for the search process to ensure that the number of resources within the queue converges to a steady state and compute explicit expressions for the mean and variance of the number of resources within the queue at steady state. We then compare the performance of the $G/M/c$ queue with that of the $G/M/\infty$ queue for an increasing number of servers. We extend the model to consider two competing targets and show that, under specific scenarios, an additional target is beneficial to the original target. Finally, we study the effects of multiple searchers. Using renewal theory, we numerically compute the inter-arrival time density for $M$ searchers in the Laplace space, which allows us to exploit the explicit expressions for the steady-state statistics of the number of resources within $G/M/1$ and $G/M/\infty$ queues, and compare their behaviour with different numbers of searchers. Overall, the $G/M/c$ queue shows a tighter dependence on the configuration of the search process than the $G/M/\infty$ queue does.
\end{abstract}

\vspace{2pc}
\noindent{\it Keywords}: search processes, first-passage times, queueing theory, renewal theory 

\submitto{\JPA}
\maketitle 


\section{Introduction}\label{sec:Intro}
Queues and random search strategies are prevalent in the natural world. Random search strategies have proven valuable when searching for one or multiple targets in an unknown domain. Examples include animal movement \cite{bartumeus_animal_2005, james_efficient_2010, bartumeus_optimal_2009}, cellular transport \cite{schwarz_optimality_2016, jandhyala_applications_2018, halford_how_2004} and software testing and optimisation \cite{zhang_random_2022, zeng_artificial_2022}. On the other hand, any system in which customers---or resources---arrive randomly and receive service provided by a set of servers at random times can be studied as a queueing process. Queues are ubiquitous in human activities, such as telecommunications \cite{alfa_queueing_2010, walraevens_stochastic_2012, massey_analysis_2002} and transportation \cite{alam_queueing_2021, gwiggner_data_2014, nakamura_queueing_2021}, and in the broader natural world, such as gene expression and regulation \cite{arazi_bridging_2004, zhang_stationary_2019, szavits-nossan_solving_2024} and metabolic pathways \cite{evstigneev_theoretical_2014, clement_stochastic_2020, kloska_queueing_2021}.

Queueing processes \cite{allen_probability_1978, asmussen_applied_2003, harchol-balter_performance_2013, bhat_introduction_2015} and stochastic search processes \cite{redner_guide_2001,Grebenkov24} have been extensively studied. More recently, they have been combined to develop a queueing model for the accumulation of resources in a target, owing to multiple rounds of search-and-capture processes \cite{bressloff_search-and-capture_2019,bressloff_queueing_2020}. The basic idea is to map the delivery of resources to customers arriving at a service station and the departure of customers from the station once they have received service, to the consumption---or degradation---of resources. So far, this mapping has always been made with the assumption that resources are consumed independently, rendering the number of servers in the service station infinite. This implies that the appropriate queueing system is a $G/M/\infty$ queue, where $G$ denotes a general inter-arrival time density, $\mathcal{F}$, $M$ denotes Markovian service times and $\infty$ specifies that the service station has an infinite number of servers and consumes resources independently. The assumption of independent consumption makes the sojourn times independent and ensures the existence of a steady state, regardless of the spatial configuration of the search process \cite{bressloff_queueing_2020, liu_gixg_1990}. 

In this study, we change the consumption protocol from independent to sequential consumption so that there are a finite number of servers in the system consuming resources in the order they arrive. The appropriate queueing system becomes $G/M/c$, where $c$ denotes the number of servers \cite{takacs_introduction_1962, cooper_introduction_1981}. In this new system, when the number of resources exceeds the number of servers, waiting times before service emerge and sojourn times cease to be independent. As a motivating example, consider the transport of proteins and vesicles along the axon of a neuron. Neurons synthesise most of their necessary molecular components in the cell body or soma, including those that ensure healthy maintenance of the axon and synaptic cell-cell communication \cite{brown_axonal_2013, hirokawa_molecular_2005}. Malfunctions in axonal transport may lead to the atypical accumulation of cellular resources in the axon and the synaptic targets and trigger neurodegenerative diseases \cite{de_vos_role_2008}. In previous work, the stochastic accumulation of resources in one or more synaptic targets due to the active transport and delivery of constitutive proteins was studied as a search-and-capture process with a $G/M/\infty$ queue \cite{bressloff_queuing_2021}. In this particular model, service in the queue was interpreted as a form of protein degradation, with no constraints on the number of proteins that could degrade over a given time interval. This is equivalent to having an infinite number of servers. On the other hand, if one is interested in the transport of neurotransmitter-rich synaptic vesicles, then a more realistic scenario is to consider the sequential utilisation or consumption of delivered vesicles during synaptic processing, which is better modelled as a $G/M/c$ queue. Here $c$ represents the maximum number of vesicles that can be processed simultaneously by fusing to the presynaptic plasma membrane, for example.

One of the major consequences of having a finite number of servers is that we must now consider scenarios in which the spatial configuration of the search process causes the resources in the system to blow up, preventing the queue from converging to a steady state. Such a phenomenon cannot occur in the case of a $G/M/\infty$ queue, and is thus the main novel feature of the current paper. For concreteness, we develop the theory by
considering a diffusive search process inside the bounded interval $[0,L]$ with a target at one or both ends of the interval. Even in this simple domain, the spatial configuration of the search processes strongly influences the behaviour of the queueing system, which translates into nontrivial analytical and numerical results for the steady-state resource statistics for the $G/M/c$. In particular, we obtain the following main results: (i) We derive steady-state existence conditions and construct critical regions in the parameter space of the search process that ensure convergence of the resulting $G/M/c$ queue to the steady state. Furthermore, under convergence to a steady state, we compute the mean and variance of the queue length. (ii) We relate the mean waiting time before service to the Wasserstein distance and utilise it to examine the convergence of the $G/M/c$ to $G/M/\infty$ as $c\to\infty$. (iii) We extend the steady state criterion to the case of a single searcher and two targets and the case of a single target with multiple searchers.

The structure of the paper is as follows. In \Sref{sec:2}, we briefly review the classical first-passage time (FPT) problem for diffusion in $[0,L]$ with a single absorbing target at $x=0$ and a reflecting boundary at $x=L$ \cite{redner_guide_2001}. We also provide an overview of queueing theory and the $G/M/c$ queue \cite{takacs_introduction_1962, cooper_introduction_1981} and formulate the accumulation of target resources in terms of said queue. In \Sref{sec:3}, we combine the theoretical frameworks from \Sref{sec:2} to derive conditions for the existence of a steady state, defining regions of convergence and blow-up. Moreover, we derive explicit expressions for the steady-state resource statistics and analyse their dependence on the parameters $x_{0}$, $L$, and $c$, where $x_0$ is the initial position of the searcher. In \Sref{sec:4}, we analyse the effects of adding servers to a $G/M/1$ at the blow-up threshold. We compare the performance of the $G/M/c$ to that of the $G/M/\infty$ using the mean waiting time before service, which is shown to be analogous to the Wasserstein distance between waiting time densities \cite{wasserstein1969markov, dobrushin_prescribing_1970}.
 
Having studied the simplest case of a single target and single searcher, we extend the model to consider two targets and multiple searchers. In \Sref{sec:5}, we analyse the scenario of two competing targets. We derive a general expression for the density of the inter-arrival times in the Laplace space, using methods that have been developed by several authors before \cite{belan_restart_2018, chechkin_random_2018, bressloff_search-and-capture_2019,bressloff_modeling_2020, bressloff_search_2020} and study the effects of unloading and loading times in the competition between both targets. Finally, in \Sref{sec:6}, we examine the effects of multiple searchers. We develop analytical results for the general $G/M/c$ queue and use numerical results to compare the steady-state resource statistics of the $G/M/1$ and the $G/M/\infty$. We find that the steady-state resource statistics of the $G/M/1$ queue grow exponentially with the number of servers, whereas those of the $G/M/\infty$ queue grow linearly.

\section{Multiple search-and-capture events and queueing theory}\label{sec:2}
\begin{figure}[h!]
	\begin{center}
		\includegraphics[width = 0.7\textwidth]{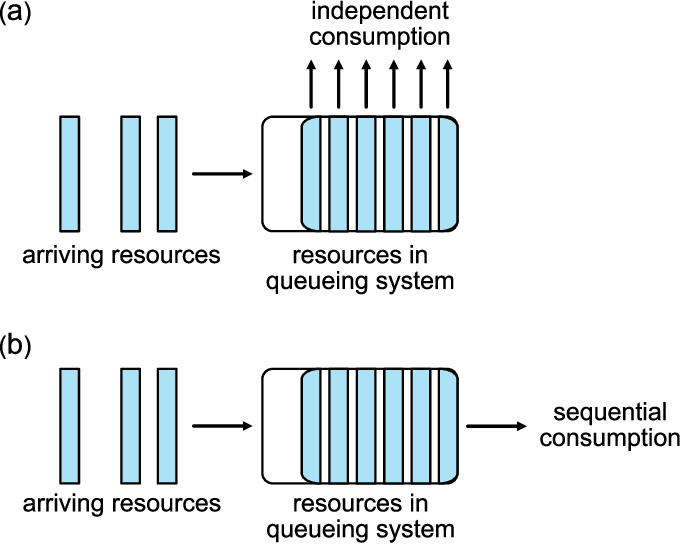}
		\caption{Different consumption protocols resulting from having an infinite and a finite number of servers. (a) An independent consumption model in which each resource within the target is consumed independently of the number of resources in the system. This consumption protocol coincides with the one of the $G/M/\infty$ queueing system. (b) A sequential consumption protocol in which resources start to be consumed in the order they are received (first-come, first-served). This consumption protocol coincides with the one of the $G/M/c$ queueing system.}
		\label{fig:ConsumptionProtocol}
	\end{center}
\end{figure}

This section presents the modelling rationale that connects search processes with queueing theory. Such a connection has previously been explored under the assumption of an independent consumption protocol \cite{bressloff_queueing_2020, bressloff_first-passage_2021, bressloff_search-and-capture_2019, bressloff_modeling_2020}, see \Fref{fig:ConsumptionProtocol}(a). However, the connection between search processes and queueing theory in the case of sequential consumption, see \Fref{fig:ConsumptionProtocol}(b), remains largely unexplored. We focus on a single particle diffusing in a finite interval with an absorbing target at one end, which sequentially consumes accumulated resources. This change in the consumption protocol is reflected in the resulting queueing model, shifting the scope of the queueing theory from the $G/M/\infty$ to a $G/M/c$, and the particular case of the $G/M/1$. Regardless of the consumption protocol, the inter-arrival time distribution is preserved among all models.

\subsection{First-passage theory}
\begin{figure}[h!]
	\begin{center}
		\includegraphics[width = 0.9\textwidth]{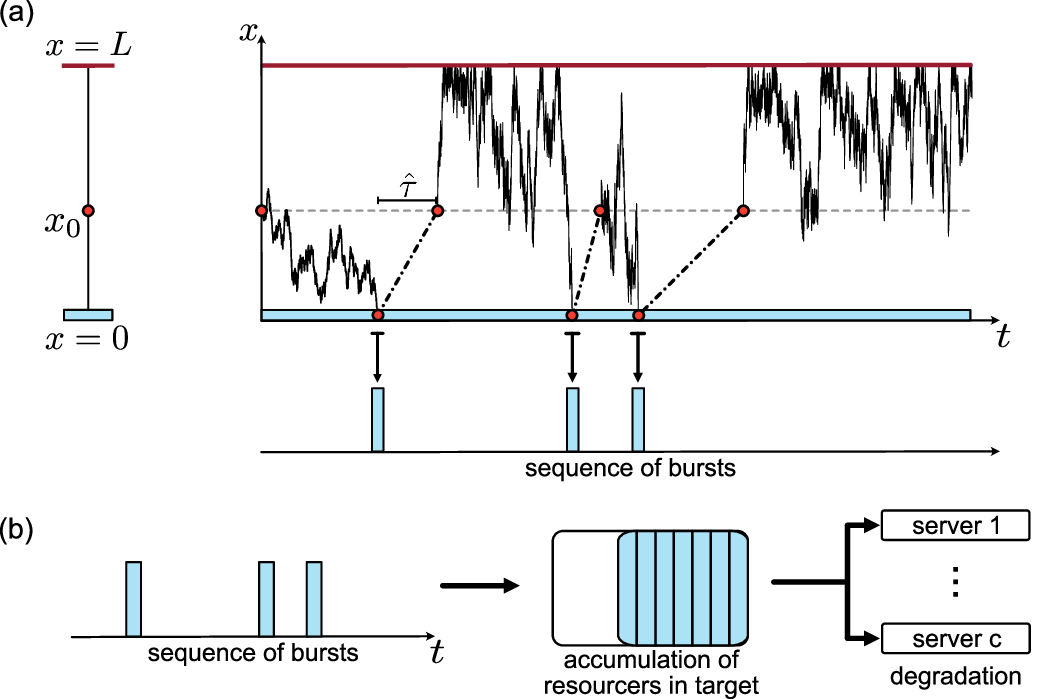}
		\caption{Continual delivery and sequential consumption of resources in a single target after several rounds of search-and-capture processes. (a) Multiple search-and-capture events for a particle diffusing inside the finite interval $[0,L]$ with an absorbing target at the origin and a reflecting boundary at $L$. Each time the particle finds the target, it delivers a resource (burst event) and returns to its initial position $x_0$, reloading and restarting the process (dash-dotted lines). The time it takes the particle to unload, return to its initial position and reload, $\hat{\tau}$, is assumed to be random. (b) The sequence of captures by the target maps into the sequence of times when the burst events occur. These times are fed into a queueing system with $c$ servers, each consuming resources within a random, exponentially distributed time. This mapping allows us to study the accumulation of resources in a target with a sequential consumption protocol as a $G/M/c$ queueing system.}
		\label{fig:modelingScheme}
	\end{center}
\end{figure}

Consider a particle following an unbiased Brownian motion in the finite interval $[0,L]$ with an absorbing boundary at $x=0$ and a reflecting boundary at $x=L$. The particle's position at time $t$, denoted by $X(t)$, evolves according to the overdamped Langevin equation
\begin{equation}\label{eq:OverdampedLangevine}
	\frac{dX(t)}{dt} = \sqrt{2D}\eta(t),
\end{equation}
where $D$ is the diffusivity and $\eta(t)$ is a Gaussian white noise with $\langle\eta(t)\rangle=0$ and $\langle\eta(t)\eta(t')\rangle=\delta(t-t')$. The absorbing boundary represents a target that accumulates resources via multiple rounds of search-and-capture. That is, each time the particle finds the target, it delivers a packet of resources (cargo) and then returns to its initial position to reload. After reloading with cargo, the particle initiates a new search-and-capture cycle, see \Fref{fig:modelingScheme}(a). This results in a random sequence of target deliveries, which we refer to as burst events. We also assume that the total time $\hat{\tau}$ for the particle to unload its resources at the target, return to its initial position and reload is a random variable with probability density $\varphi(\hat{\tau})$ and mean $\tauc$. We will refer to $\hat{\tau}$ as the total refractory time.

We denote the arrival time of the $n$th burst as $\tau_n$. Observe that, once the search-and-capture process has started, the inter-arrival time between any two bursts follows the same distribution, given by the sum of the total refractory time and the FPT to the target, having started at $x_{0}$, defined as
\begin{equation}
	\mathcal{T}(x_{0}) := \inf\{t\geq 0 : X(t)=0,\,X(0)=x_{0}\},
\end{equation}
with the convention $\inf\{\emptyset\}=\infty$; $\mathcal{T}(x_{0})=\infty$ meaning that the particle never reaches the target. Then, the $n$th inter-arrival time, denoted by $\Delta_{n}$, can be expressed as
\begin{equation}\label{eq:InterArrivalTime}
	\Delta_{n} := \tau_{n} - \tau_{n-1} \stackrel{d}{=} \hat{\tau} + \mathcal{T}(x_{0}),
\end{equation}
where $\Delta_{n}\stackrel{d}{=}\hat{\tau} + \mathcal{T}(x_{0})$ denotes that $\Delta_{n}$ has the same distribution as $\hat{\tau} + \mathcal{T}(x_{0})$. Therefore, as the inter-arrival times are independent, identically distributed random variables (i.i.d.r.v.), the sequence of bursts can be studied as a renewal process \cite{bladt_renewal_2017}, characterised by the inter-arrival time distribution. To obtain the distribution of $\Delta_{n}$, we begin by studying the distribution of the FPT. 

The probability density for the particle to be in $x$ at time $t$, having started at $x_{0}$, is denoted by $p(x,t|x_{0})$. As the particle follows pure Brownian motion, governed by the overdamped Langevin equation \eref{eq:OverdampedLangevine} away from the boundaries, its probability density function evolves according to the diffusion equation
\begin{equation}\label{eq:diffusion}
	\frac{\partial p(x,t|x_{0})}{\partial t} = D\frac{\partial^{2}p(x,t|x_{0})}{\partial x^{2}} = -\frac{\partial J(x,t|x_{0})}{\partial x},
\end{equation}
where $D$ is the diffusivity and $J(x,t|x_{0}) := -D\partial_{x}p(x,t|x_{0})$ is the probability flux \cite{Gardiner_stochastic_1985}. The diffusion equation is supplemented by the absorbing boundary condition $p(0,t|x_{0})=0$, the reflecting boundary condition $\partial_{n}p(L,t|x_{0})=0$ and the initial condition $p(x,0|x_{0})=\delta(x-x_{0})$. 

As only one target exists, the FPT density $f_{0}(t)$ for a fixed $x_0$ is equivalent to the probability flux into the absorbing boundary at $x=0$. Thus, we obtain the following expression for the Laplace transform of the FPT density \cite{redner_guide_2001}:
\begin{equation}\label{eq:de}
	\widetilde{f}_{0}(s) = D\frac{\partial}{\partial x}\widetilde{p}(x,t|x_{0})\Big|_{x=0} = \frac{\cosh\left[\sqrt{\frac{s}{D}}(L-x_{0})\right]}{\cosh\left[\sqrt{\frac{s}{D}}L\right]},
\end{equation}
where $\widetilde{f}_{0}(s) := \int_{0}^{\infty}e^{-st}f_0(t)dt$ is the Laplace transform of the function $f_{0}$. The mean first-passage time (MFPT) can be obtained directly as
\begin{equation}\label{eq:MFPT}
	T(x_{0}) = -\frac{d}{d s}\widetilde{f}_{0}(s)\Big|_{s=0} = \frac{(2L -x_0)x_0}{2D}.
\end{equation}
Finally, the density of the inter-arrival times is given by the convolution of the FPT density and the refractory time density,
\begin{equation}\label{eq:inter-arrival-density}
	\mathcal{F}(t) = (f_{0}*\varphi)(t) := \int_{0}^{t}f_{0}(t')\varphi(t-t')dt'.
\end{equation}
Laplace transforming \eref{eq:inter-arrival-density} yields
\begin{equation}\label{eq:inter-arrival-laplace}
	\widetilde{\mathcal{F}}(s) = \widetilde{f}_{0}(s)\widetilde{\varphi}(s).
\end{equation}
Given all of the above computations, the mean inter-arrival time can be determined as $T(x_{0}) + \tauc$, and the mean rate of resource delivery to the target is $\lambda = (T(x_{0}) + \tauc)^{-1}$. Having determined the inter-arrival time density and mean inter-arrival time, we focus on the queueing theory and its connection with search processes.
\subsection{Queueing theory}
To track the accumulation of resources within the target after several deliveries from the search-and-capture process, we resort to classical queueing theory. Referring to the elements in a queueing system as customers is common, so we will adopt this terminology. To describe a queueing system analytically, one needs to determine the customer's arrival distribution, number of servers, queue discipline and maximum queueing system capacity \cite{allen_probability_1978}. The usual scenario assumes a system with infinite capacity and a first-come, first-served (FCFS) discipline. In this scenario, we can define the queue using Kendall's shorthand notation $A/B/c$, where $A$ describes the inter-arrival time distribution, $B$ the service time distribution and $c$ the number of servers \cite{kendall_stochastic_1953}.

The resources delivered during the burst events are considered to be the customers in the queueing system, making the inter-arrival time distribution a general distribution. The resource consumption, occurring at rate $\mu$, is equivalent to the customers leaving the system after receiving service. Therefore, the service times are assumed to be exponentially distributed i.i.d.r.v.\ with the same rate parameter $\mu$. This indicates that the corresponding queueing system must have general, independent inter-arrival times, represented by the symbol $G$---or in some cases $GI$---,and Markovian service times, represented by the symbol $M$. Regarding the number of servers, previous works assumed that resources are consumed independently from each other, forcing the system to have an infinite number of servers and making the $G/M/\infty$ system the appropriate choice \cite{bressloff_queueing_2020, bressloff_first-passage_2021, bressloff_search-and-capture_2019, bressloff_modeling_2020}. In this study we consider the resources to be consumed sequentially, following a FCFS discipline in a system with a finite number of servers. This means that the $G/M/c$ queue is now an appropriate system for studying the accumulation of resources in the target, see \Fref{fig:modelingScheme}(b). As the $G/M/c$ system has been extensively studied, we will limit ourselves to stating the main results regarding its limit steady-state distribution, through which we will determine the steady-state statistics of resource accumulation in the target. For further details on the steady-state distribution of the $G/M/c$, refer to Refs. \cite{cooper_introduction_1981, takacs_introduction_1962} and \ref{ap:GMC}, and for further details on the steady-state distribution of the $G/M/\infty$ see Refs. \cite{liu_gixg_1990, takacs_coincidence_1958}.

\begin{figure}[h!]
	\begin{center}
		\includegraphics[width = \textwidth]{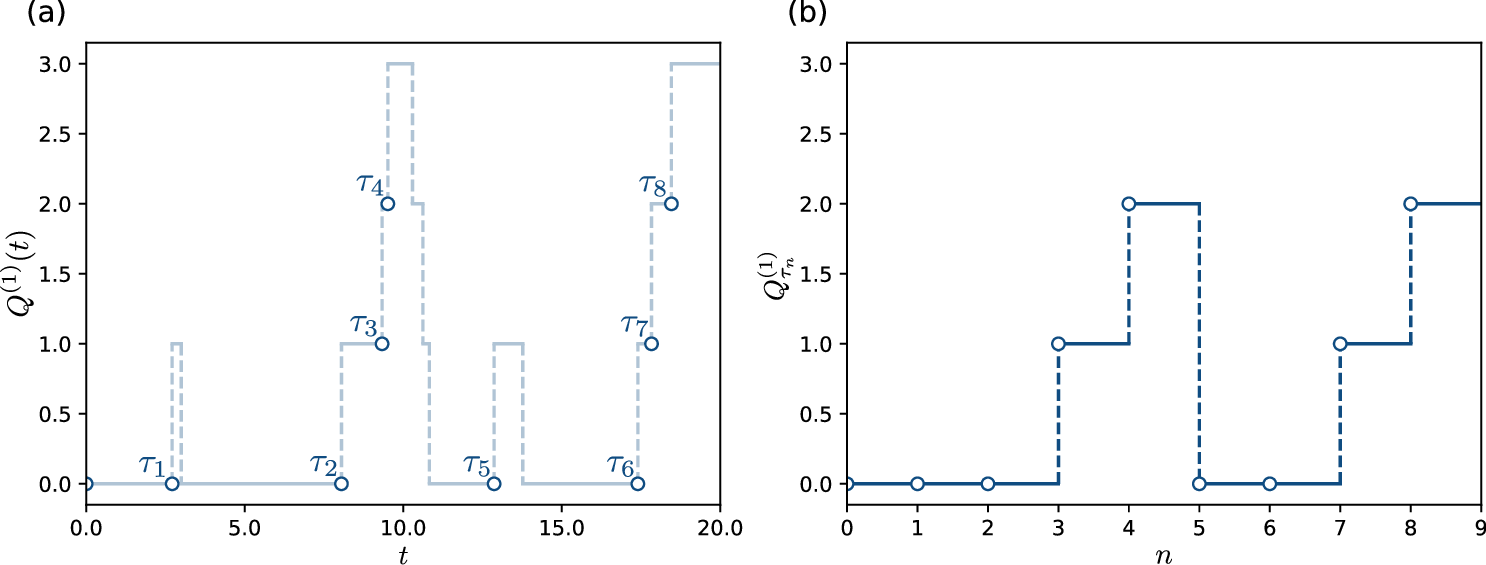}
		\caption{Sample path of the queue length of a $G/M/1$ system, $Q^{(1)}(t)$, and its embedded Markov chain, $Q^{(1)}_{\tau_{n}}$. (a) Sample path of the queue length of a $G/M/1$ system. The first eight arrival times are represented by $\{\tau_{1},\dots,\tau_{8}\}$ and the empty dots represent the number of customers in the system just prior to each one of these arrivals. (b) Sample path of the embedded Markov chain. The steps in this discrete-time embedded Markov chain correspond to the arrival times in the whole system and the value it takes in each step corresponds to the number of customers in the system just prior to the respective arrival time, again represented by empty dots.}
		\label{fig:EMC}
	\end{center}
\end{figure}

Let $Q^{(c)}(t)$ be the number of customers in the system at time $t$, $\tau_{n}$ be the time of arrival of the $n$th customer and let $Q^{(c)}_{\tau_{n}}=Q^{(c)}(\tau_{n}^-)$, $\tau_n^-=\lim_{\epsilon \rightarrow 0^+}\tau_n-\epsilon$, be the number of customers waiting in the system with $c$ servers ahead of the $n$th customer just prior to the time of their arrival. $Q^{(c)}_{\tau_{n}}$ defines a discrete-time embedded Markov chain (see \ref{ap:GMC}). Note that this discrete-time Markov chain jumps each time a new customer arrives at the queueing system, and it takes the value of the number of customers the arriving customer finds ahead of them in the system, see \Fref{fig:EMC}. Therefore, its steady-state distribution is often called the \textit{arriving customer distribution}. It can be shown that there exists a unique steady-state arriving customer distribution 
\begin{equation}\label{eq:EmMC_limit}
	\Pi_j^{(c)} = \lim_{n\to\infty} \mathbb{P}[Q^{(c)}_{\tau_{n}} = j],
\end{equation}
if and only if 
\begin{equation}\label{eq:rho-leq1}
\rho=\lambda/(c\mu)<1,
\end{equation}
where $\rho$ is the \textit{traffic intensity} or \textit{server occupancy} \cite{takacs_introduction_1962}, and $\lambda$ is the mean customer arrival rate. For the proof, see \ref{ap:GMC}. Observe that $\lim_{c\to\infty}\rho = 0$, coinciding with the $G/M/\infty$ queue always having a steady-state distribution. 

Although in this work we will only deal with $G/M/c$, it is worth mentioning that the condition for convergence to the stationary distribution in terms of traffic intensity is the same regardless of whether the inter-arrival time distribution and the service time distribution are Markovian. This is to say, if we consider the most general $G/G/c$ queueing system, with mean customer arrival rate $\lambda$ and mean service rate $\mu$, then its queue length converges to a steady-state if and only if $\rho=\lambda/(c\mu)<1$. The interpretation is as follows. Suppose that the system is working at full capacity, i.e. with all of its servers busy, for a large amount of time $t$. By the Law of Large Numbers (LLN), there will be about $t\lambda$ arrivals and about $tc\mu$ services ($t\mu$ for each server). Thus $\rho$ is about the ratio of arrivals and services and when $\rho>1$ the number of arrivals exceeds the number of services, so we expect the queue to grow indefinitely \cite{asmussen_applied_2003}. Note that this condition relies solely on the long-term average workload balance and not distributional details, e.g. Markovianity.

The arriving customer distribution can be explicitly determined. However, determining this distribution requires significant labour, and the resulting formulas do not shed any light on this problem. Here, we state the resulting distribution only for the $G/M/1$ and provide further details in \ref{ap:GMC}. In the particular case of the $G/M/1$ system, the steady-state arriving customer distribution boils down to a geometric distribution with parameter $1-\sigma^*$
\begin{equation}\label{eq:GM1_stationary}
	\Pi_j^{(1)} = (1-\sigma^*)(\sigma^*)^{j},\qquad j\geq 0,
\end{equation}
where $\sigma^*$ is the solution to the following implicit equation
\begin{equation}\label{eq:sigma_star}
	 \Phi(\sigma) := \widetilde{\mathcal{F}}(c\mu(1-\sigma)) = \sigma.
\end{equation}
It is worth noting that for Poisson arrivals with rate $\lambda$, e.g. a $M/G/c$ system, the solution $\sigma^{*}$ of  \eref{eq:sigma_star} equals the traffic intensity ($\sigma^{*}=\rho$); for this reason, $\sigma^{*}$ is sometimes known as the \textit{generalised traffic intensity} \cite{cooper_introduction_1981}. However, in general, the generalised traffic intensity for a $G/M/c$ queue is not the same as the true traffic intensity ($\sigma^{*}\neq\rho$). For example, a $G/M/1$ system with arrival distribution determined by a search-and-capture process in the interval $[0,3]$ with starting position $x_{0}=1$, has a traffic intensity of $0.4$ and a generalised traffic intensity of approximately $0.5258$. In addition, given the lack of Markovianity in the inter-arrival time distribution, the steady-state arriving customer distribution depends on the entire inter-arrival distribution via $\sigma^{*}$, as it can be seen in \eref{eq:GM1_stationary} for the $G/M/1$ and \eref{eq:Ui_AP}–\eref{eq:Probs_Fin_AP} for the $G/M/c$.

\begin{figure}[h!]
	\begin{center}
		\includegraphics[width = 0.65\textwidth]{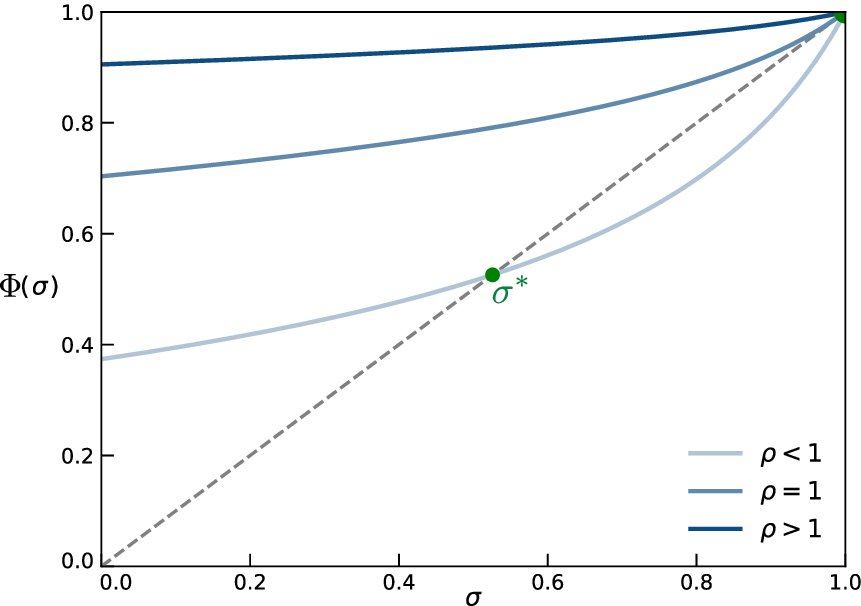}
		\caption{Analysis of \eref{eq:sigma_star} for different traffic intensities and graphical construction of solutions for $\rho<1$. When $\rho\geq1$ the only solution to $\Phi(\sigma)=\sigma$ is $\sigma=1$. In contrast, when $\rho<1$ there exists a unique $\sigma^{*}\in(0,1)$ such that $\Phi(\sigma^{*})=\sigma^{*}$.}
		\label{fig:SS}
	\end{center}
\end{figure}

Before proceeding, we develop a graphical construction of the solutions to \eref{eq:sigma_star}. First, let us remember the following relationship between the Laplace transform of the inter-arrival times density and the moment-generating function of the inter-arrival times
\begin{equation}\label{eq:Laplace-moments}
	 \widetilde{\mathcal{F}}(s) = M_{\Delta_{n}}(-s) = \mathbb{E}[e^{-s\Delta_{n}}].
\end{equation}
From this relation, since $\Phi(1)= \mathbb{E}[e^{0}]=1$, it follows that $\sigma=1$ is always a solution to \eref{eq:sigma_star}, regardless of the values of $c, \mu$ and $\lambda$. In addition, $\Phi(0) = \mathbb{E}[e^{-c\mu\Delta_{n}}]>0$. It is straightforward to show that
\begin{equation}\label{eq:Phi_prime}
	 \Phi'(\sigma) = c\mu\mathbb{E}[\Delta_{n}e^{-c\mu(1-\sigma)\Delta_{n}}] > 0,
\end{equation}
as $\Delta_{n}$ and $e^{-c\mu(1-\sigma)\Delta_{n}}$ are nonnegative random variables and
\begin{equation}\label{eq:Phi_prime_1}
	 \Phi'(1) = c\mu\mathbb{E}[\Delta_{n}] = \rho^{-1}.
\end{equation}
Finally,\begin{equation}\label{eq:Phi_double_prime}
	 \Phi''(\sigma) = (c\mu)^{2}\mathbb{E}[\Delta_{n}^{2}e^{-c\mu(1-\sigma)\Delta_{n}}] > 0,
\end{equation}
as, again, $\Delta_{n}^{2}$ and $e^{-c\mu(1-\sigma)\Delta_{n}}$ are nonnegative random variables. In summary, $\Phi(\sigma)$ is a positive-definite, convex, monotonically increasing function of $\sigma$ for $\sigma\in[0,1]$. Furthermore, if $\Phi'(1)>1$ then the graphical construction in \Fref{fig:SS} establishes the existence of a unique solution $\sigma^{*}\in(0,1)$ that satisfies \eref{eq:sigma_star}. Observe that $\Phi'(1)>1$ only when $\rho<1$.

There is a crucial distinction between the arriving customer distribution obtained for the embedded Markov chain and the steady-state distribution of the original queueing system that evolves in continuous time. This original steady-state distribution for the queue length is often known as the \textit{outside observer distribution}. Let us denote the outside observer distribution for the $G/M/c$ by $\{P^{(c)}_j\}_{j\geq0}$, where $P^{(c)}_{j}=\lim_{t\to\infty} \mathbb{P}[Q^{(c)}(t) = j]$. Observe that, for each $j\geq0$, $\Pi^{(c)}_{j}=\lim_{n\to\infty} \mathbb{P}[Q^{(c)}_{\tau_{n}} = j]$ is the probability that an arbitrary arriving customer finds $j$ customers in the system ahead of them. In contrast, for each $j\geq0$, $P^{(c)}_{j}$ is the probability that an outside observer finds $j$ customers in the system at any given time once the system has reached the steady state. For a clarifying example, see \ref{ap:ArrVsOut}. In general, these two distributions are different. However, it can be shown that
\begin{equation}\label{eq:Outside_GMC}
	P_j^{(c)} = \frac{\lambda}{\mu(j)}\Pi_{j-1}^{(c)},\qquad j\geq 1,
\end{equation}
where $\mu(j) = j\mu$ if $j<c$ and $\mu(j) = c\mu$ if $j\geq c$ \cite{takacs_introduction_1962, heyman_relation_1980}. Note that \eref{eq:Outside_GMC} defines only the outside observer distribution for $j\geq1$. However, the steady-state probability of having an empty system can be obtained as
\begin{equation}\label{eq:Outside_GMC_P0}
	P_0^{(c)} =1-\sum_{j=1}^{\infty}P_j^{(c)} .
\end{equation}

We present some of the commonly defined intrinsic times of a queueing system and well-known expressions for their distributions. Once a customer enters the queueing system, three relevant times are associated with their entire service. First, there is the time they wait in the queue before they can reach a server and start their service. Once the customer reaches the server, they must wait until their service is complete. The last relevant time is the time the customer spends in the system, known as the \textit{sojourn time}. Note that the sojourn time is always the same as the sum of the waiting time before service and service time. With this in mind, let us denote by $W$ the time an arbitrary customer waits for their service to begin once the system reaches a steady state. It can be shown that 
\begin{equation}\label{eq:WaitingTimeSurvival}
	\mathbb{P}[W>t] = \frac{A}{1-\sigma^*}e^{-(1-\sigma^*)c\mu t},
\end{equation}
where $A$ is a normalisation constant for the steady-state distribution $\Pi_j^{(c)}$ of the $G/M/c$ queue. In the case $c=1$, we define $A$ according to $\Pi_j^{(c)} =A (\sigma^*)^{j-1}$ such that $\sum_{j=0}^{\infty}\Pi_j^{(c)}=1$, which implies $A=\sigma^{*}(1-\sigma^{*})$. The definition and calculation of $A$ for $c>1$ is more involved, see \eref{eq:A_AP} of \ref{subap:A3}. For the derivation of \eref{eq:WaitingTimeSurvival}, see \ref{ap:WaitingTime}. As the waiting time is a nonnegative random variable, we have
\begin{equation}\label{eq:WaitingTimeE}
	\mathbb{E}[W] = \int_0^\infty \mathbb{P}[W>t]d t = \frac{A}{(1-\sigma^*)^2c\mu}.
\end{equation}
In the special case of the $G/M/1$ \eref{eq:WaitingTimeSurvival} and \eref{eq:WaitingTimeE} take the following form
\begin{equation}\label{eq:WaitingTimeEGM1}
	\mathbb{P}[W>t] = \sigma^*e^{-(1-\sigma^*)\mu t}, \qquad \mathbb{E}[W] = \frac{\sigma^*}{\mu(1-\sigma^*)}.
\end{equation}
Denoting the service time for an arbitrary customer by $T$ with $T\sim \exp(\mu)$, the corresponding sojourn time $S$ is
\begin{equation}\label{eq:SojournTime}
	S = W + T,
\end{equation}
with the following mean
\begin{equation}\label{eq:MeanSojourn}
	\mathbb{E}[S] = \frac{A + (1+\sigma^*)^2c}{(1-\sigma^*)^2c\mu}.
\end{equation}
In the particular case of the $G/M/1$, the mean sojourn time is given by
\begin{equation}\label{eq:MeanSojournGM1}
	\mathbb{E}[S] = \frac{1}{(1-\sigma^*)\mu}.
\end{equation}

Note how the steady-state distribution, the waiting time and the sojourn time in the $G/M/c$ system all depend on the entire inter-arrival time distribution via $\sigma^{*}$ and $A$. This is because the Markovianity of the service times shifts the complexity of the analysis to the arrival process and its distribution. This dependence on the entire distribution contrasts with the results for models with Poisson arrivals, $M/G/c$, in which performance measures, such as mean waiting time and mean queue length, depend only on the mean and variance of the waiting times, despite their lack of Markovianity \cite{cooper_introduction_1981}. This is due to the unique property of Poisson arrivals that the fraction of arriving customers that find the system in some state $E$ is the same as the fraction of time an outside observer finds the system in state $E$. This property is called PASTA (Poisson Arrivals See Time Averages) \cite{bhat_introduction_2015}. Therefore, the PASTA property allows one to directly determine mean performance measures of the $M/G/c$ in terms of the mean and variance of the waiting times \cite{adan_mean_2011}.
\section{Blow-up conditions and steady-state statistics}\label{sec:3}
We can now combine the results from \Sref{sec:2} to determine the conditions under which the system does not blow up, and analyse the number of resources in the system when it reaches a steady state. Furthermore, we analyse the effects of increasing the number of servers in terms of the blow-up regions and steady-state statistics for the $G/M/c$. Hereafter, we fix the units of time and length by setting the consumption rate $\mu=1$ and diffusivity $D=1$. We begin our analysis by assuming instantaneous unloading, return to $x_0$ and loading. However, we later explore some of the effects of a nonzero refractory time $\hat{\tau}$.

Suppose that once the searcher has found the target, it immediately unloads the resources, returns to the initial position, loads a new cargo and starts the new search-and-capture round; this is $\varphi(\hat{\tau}) = \delta(\hat{\tau})$ and $\tauc=0$. Substituting the explicit expression for the MFPT \eref{eq:MFPT} into the arrival rate, we obtain
\begin{equation}\label{eq:lambda}
	\lambda = \frac{1}{T(x_0)} = \frac{2D}{(2L -x_0)x_0}.
\end{equation}
Therefore, the condition for the existence of a stationary distribution of the $G/M/c$ system in terms of the traffic intensity \eref{eq:rho-leq1} becomes
\begin{equation}\label{eq:ExistenceCondition}
	\rho =  \frac{2D}{c\mu(2L -x_0)x_0} < 1.
\end{equation}
From this condition, we obtain new threshold quantities for the interval length and initial position to ensure convergence to a steady-state distribution. First, given a fixed interval length $L$, we find a threshold value for the initial position, $x_{0}^{*}$, for which if $x_{0}\leq x_{0}^{*}$ then the queueing system blows up, and if $x_{0}>x_{0}^{*}$ the system converges to a steady state. The critical point for blow up is determined by setting $\rho = 1$. This yields a quadratic equation for $x_{0}$:
\begin{equation}\label{eq:ExistenceThresholdCuadratic}
	x_0^2 - 2Lx_{0}+\frac{2D}{c\mu}=0.
\end{equation}
Suppose that we fix $c$ and take $L$ to be sufficiently large so
that the discriminant $4[L^2- 2D/(c\mu)]>0$, that is,
\begin{equation*}
 L\geq\sqrt{\frac{2D}{c\mu}},
\end{equation*}
There then exists a pair of positive real roots, one of which lies within the physical domain $[0,L]$, namely,
\begin{equation}\label{eq:ExistenceThreshold}
x_0=x_0^*(L) \equiv  L-\sqrt{L^2-\frac{2D}{c\mu}}.
\end{equation}
We find that $\rho>1$ for all $0<x_0<x_0^*(L)$ and $\rho<1$ for $x_0^*(L)<x_0<L$. 
Define $L^{*}$ as the length for which the discriminant vanishes,
\begin{equation}\label{eq:CritialL}
L^* = \sqrt{\frac{2D}{c\mu}},
\end{equation}
It follows that if $L<L^*$ then \eref{eq:ExistenceThresholdCuadratic} has no real roots and  $\rho>1$ regardless of the starting position $x_{0}$. Finally, in the special case $L=L^*$, we have $x_{0}^{*}=L^{*}$ and any initial position makes the system blow up.

\begin{figure}[h!]
	\begin{center}
		\includegraphics[width = 0.65\textwidth]{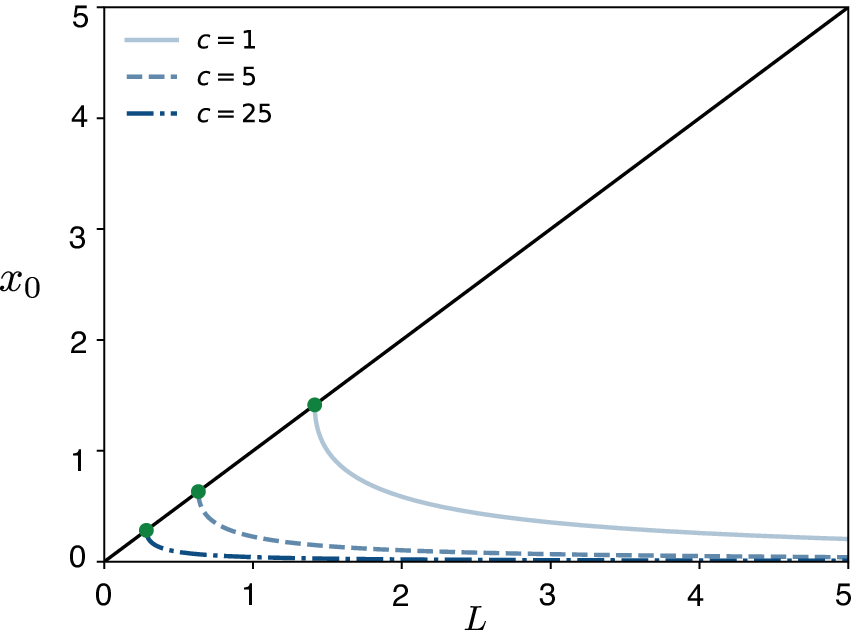}
		\caption{Critical spatial configurations for the existence of a steady-state distribution. Each curve represents the threshold spatial configurations for the search-and-capture process that make the $G/M/c$ queueing system for the accumulation of resources blow up. These curves correspond to \eref{eq:ExistenceThreshold} with 1, 5 and 25 servers in a $G/M/c$ queueing system. Any pair of values $(L,x_{0})$ below the diagonal line and above the curve corresponding to a value of $c$ defines a spatial configuration that allows the queueing system with said number of servers to relax to a steady state. The green points represent the maximum interval length for which a steady-state distribution never exists in each case. In every case, we fix $D=1$ and $\mu=1$.}
		\label{fig:CriticalLines}
	\end{center}
\end{figure}

Therefore, to have a spatial configuration that allows the system to converge to a steady state, the interval length must fulfil $L>L^*$, and the initial position must fulfil $x_{0}>x_0^*$, see \Fref{fig:CriticalLines}. Note that $L^{*}$ and $x_{0}^{*}$ explicitly depend on the number of servers, $c$, but we drop the explicit dependence from the notation for the sake of simplicity. Both $L^*$ and $x_0^*$ vanish in the limit $c\to\infty$, again recovering the fact that a system with an independent consumption protocol, modelled by a $G/M/\infty$, has a steady-state distribution regardless of its spatial configuration. In addition, as the interval length increases, the particle can take longer explorations of the space before finding the target, causing a reduction in the arrival rate and critical initial position. This result agrees with the fact that $x_{0}^{*}$ vanishes in the limit $L\to\infty$.

Having determined the explicit conditions that ensure convergence to a steady state, we analyse the steady-state statistics for the accumulation of resources and compare them with those of the $G/M/\infty$ \cite{bressloff_queueing_2020}. Again, the difference in complexity of the resulting stationary distributions between the $G/M/1$ and $G/M/c$ with $c>1$ motivates separate analyses for these two cases. In the case of the $G/M/1$, the arriving customer stationary distribution is a geometric distribution with parameter $1-\sigma^*$, see \eref{eq:GM1_stationary}, allowing us to directly calculate the first and second moment from this explicit expression for the stationary distribution. (For the complete calculations, see \ref{ap:Statistics}.) For the sake of simplicity, let us define a random variable $Q_\infty^{(1)}$ representing the length of the $G/M/1$ queue at steady state. We then have that $Q_\infty^{(1)}\sim\{P^{(1)}_j\}_{j\geq0}$ and
\begin{equation}\label{eq:GM1_Moments}
	\mathbb{E}[Q_\infty^{(1)}] = \frac{1}{T(x_0)\mu(1-\sigma^*)},\qquad \mathbb{E}[(Q_\infty^{(1)})^2] = \frac{1+\sigma^*}{T(x_0)\mu(1-\sigma^*)^2}.
\end{equation}
From these two equations, we obtain the following explicit expression for the variance of the queue length at steady state
\begin{equation}\label{eq:GM1_Var}
	\mathrm{Var}(Q_\infty^{(1)})= \frac{1}{T(x_{0})\mu(1-\sigma^{*})}\left[\frac{1+\sigma^{*}}{1-\sigma^{*}} - \frac{1}{T(x_{0})\mu(1-\sigma^{*})}\right].
\end{equation}
The expression for the mean queue length at steady state coincides with Little's law \cite{little_proof_1961}. Little's law states that the mean queue length at steady state equals the product of the mean arrival rate and mean sojourn time \cite{little_littles_2011}. Therefore, we have 
\begin{equation}\label{eq:GM1_LittleLaw}
	\lambda\mathbb{E}[S] = \frac{1}{T(x_0)\mu(1-\sigma^*)} = \mathbb{E}[Q_\infty^{(1)}].
\end{equation}

As mentioned above, the stationary distribution of the queue length in the $G/M/c$ system is not as analytically tractable as in the $G/M/1$, posing several challenges for calculating the steady-state statistics of this system. Nevertheless, we invoke Stidham's version of Little's law to obtain an explicit expression for the first moment. Stidham's version of Little's law states that if the customer's arrival rate, $\lambda$, and mean sojourn time, $\mathbb{E}[S]$, exist and are finite, then the mean queue length equals the product of both \cite{stidham_technical_1974}. Therefore, assuming condition \eref{eq:ExistenceCondition} is satisfied, the $G/M/c$ queue satisfies both of the necessary conditions for Stidham's version of Little's law. We can then write the following explicit formula for the mean number of resources in the $G/M/c$ at steady state
\begin{equation}\label{eq:GMc_Moments}
	\mathbb{E}[Q_\infty^{(c)}] = \frac{A+(1+\sigma^*)^2c}{T(x_0)(1-\sigma^*)^2c\mu},
\end{equation}
where $A$ is the normalising constant defined in \eref{eq:A_AP}. There are no explicit formulas for the higher moments of the number of resources in the $G/M/c$ at steady state; therefore we resort to numerical approximations. These numerical approximations rely on the fact that the steady-state outside observer distribution of the $G/M/c$ queueing system has a rapidly decaying tail. Hence, we can approximate the higher moments by directly computing a large number of terms in the defining sum for discrete moments. 

We found that all of the steady-state outside observer distribution can be computed up to the smallest representable float larger than zero with 64-bit floating-point numbers, i.e. $5\times10^{-324}$, within a finite number of terms. Therefore, we compute the whole distribution up to the $N$th term such that $P^{(c)}_{N}\geq5\times10^{-324}$ and $P^{(c)}_{n}=0$ for $n>N$ under the 64-bit floating-point representation. Hence, this is the best approximation one can make with 64-bit floating-point numbers. The number of terms, $N$, depends on the number of servers $(c)$, the interval length $(L)$ and the starting position $(x_{0})$. However, all of the approximations carried out in this work achieved the smallest value under the 64-bit floating-point representation, $5\times10^{-324}$, with $N\leq10^{7}$. The function that computes the steady-state outside observer distribution was optimised to start the approximation with $N=10^{4}$, check if $P^{(c)}_{N}\geq5\times10^{-324}$ and if this condition is not satisfied repeat the process with $\hat{N}=N\times10$. As mentioned before, regardless of the set of parameters, this function always computes the whole representable distribution under the 64-bit floating-point representation for $N\leq10^{7}$.

\begin{figure}[h!]
	\begin{center}
		\includegraphics[width = 0.503\textwidth]{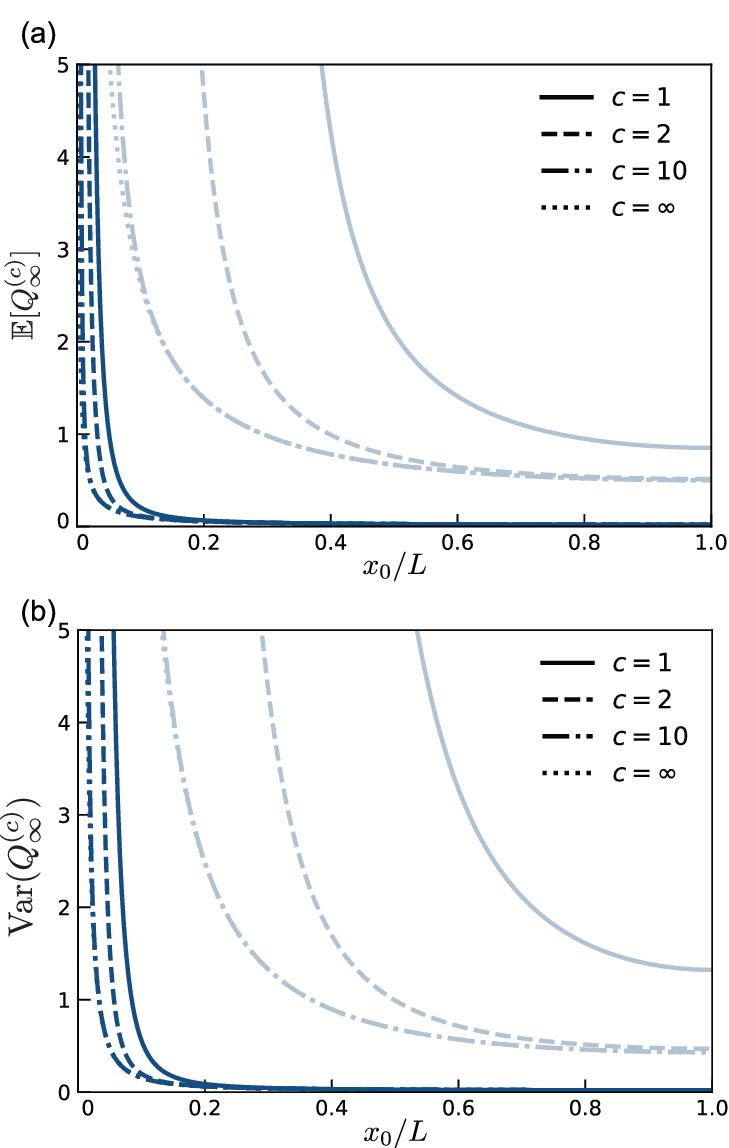}
		\caption{Mean and variance of the queue length at steady state for different number of servers. (a) Plots of the steady-state mean, $\mathbb{E}[Q_\infty^{(c)}]$, as a function of the initial position for various numbers of servers and interval lengths. (b) Plots of the steady-state variance, Var$(Q_\infty^{(c)})$, as a function of the initial position for various numbers of servers and interval lengths. In both panels, the light curves represent a system with an interval length of $L=2$, and the dark curves represent a system with an interval length of $L=10$. In every case, we fix $D=1$ and $\mu=1$.}
		\label{fig:MeanVar}
	\end{center}
\end{figure}

\Fref{fig:MeanVar}(a) shows plots of the mean number of resources in the system at steady state as a function of the initial position relative to the interval length for various numbers of servers and two different interval lengths. The corresponding plots for variance are shown in \Fref{fig:MeanVar}(b). As expected, the blow-up region for the initial position, relative to the interval length, decreases as the length, $L$, increases. This reduction in the blow-up region can be thought of as the fraction of the interval closest to $x=0$, where an initial position would cause the accumulation of resources to blow up, getting smaller as $L$ increases; i.e.\ for larger lengths, the initial position can be almost anywhere in the interval and still have convergence to a steady state. This is consistent with \eref{eq:ExistenceThreshold}, where $x_{0}^{*}=0$ as $L\to\infty$. Moreover, \Fref{fig:MeanVar} shows the rapid convergence of the steady-state statistics of the $G/M/c$ to those of the $G/M/\infty$. It can be shown that $Ac^{-1}= 0$ and $\sigma=0$ as $c\to\infty$. This makes the convergence consistent with \eref{eq:GMc_Moments} as
\begin{equation}\label{eq:GMc_GMInfty_Moments}
	\lim_{c\to\infty}\mathbb{E}[Q_\infty^{(c)}] = \frac{1}{T(x_0)\mu} = \mathbb{E}[Q_\infty^{(\infty)}],
\end{equation}
coinciding again with Little's Law and the results in \cite{bressloff_queueing_2020}.

\section{A measure of performance for the $G/M/c$ compared with $G/M/\infty$}\label{sec:4}
\begin{figure}[h!]
	\begin{center}
		\includegraphics[width =0.92\textwidth]{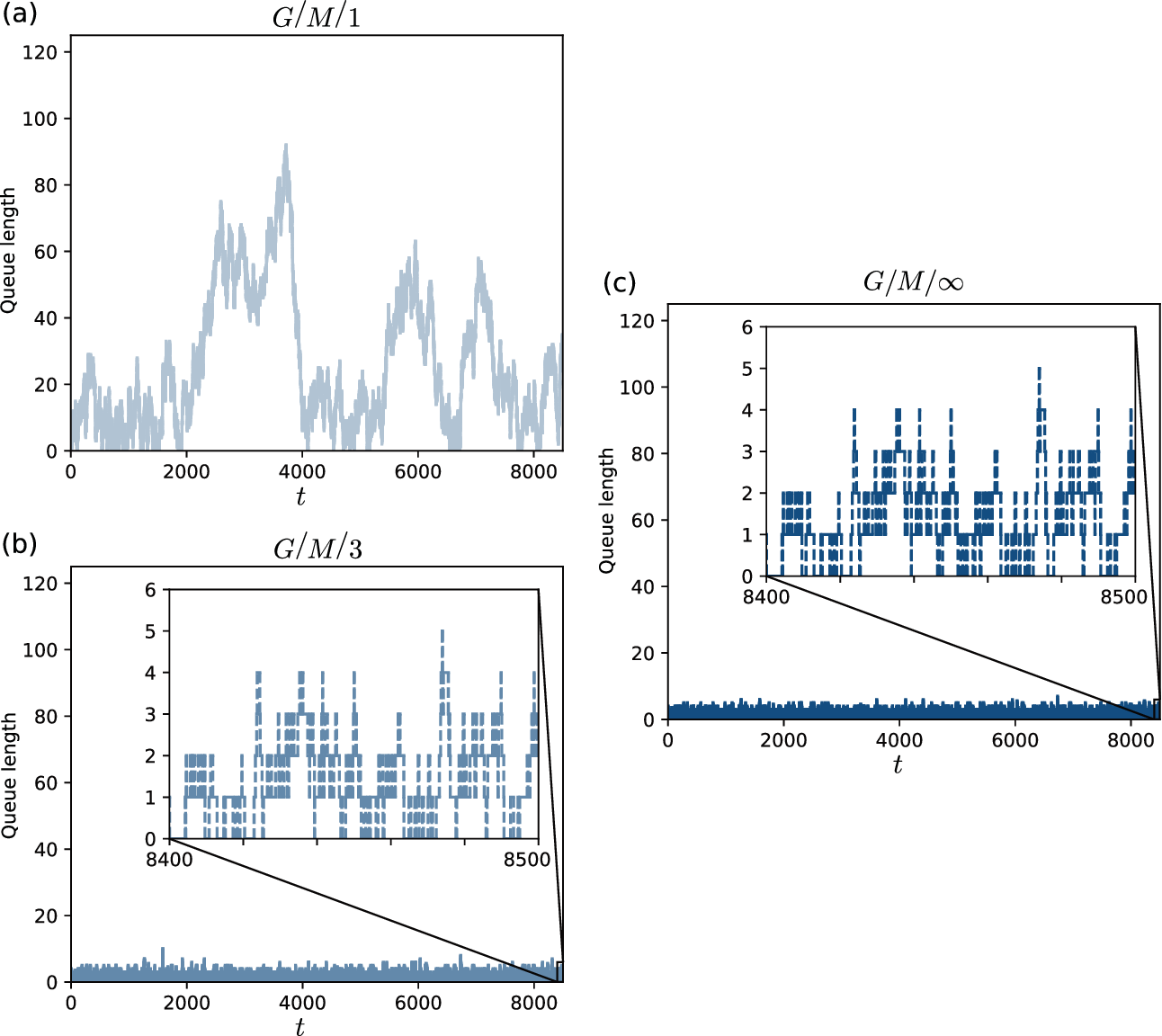}
		\caption{Sample paths of for the $G/M/1$ (a), $G/M/3$ (b) and $G/M/\infty$ (c). All simulations were carried out with $L=1.5$ and $x_{0}=1$. Note that this spatial configuration lies exactly in the critical zone where $G/M/1$ blows up, as $x_{0}^{*}=1$ when $L=1.5$, see \Fref{fig:CriticalLines}. The simulations were carried out using the same sample of arrival times and service times in the three cases, just changing the number of servers for each panel. In every case, we fix $D=1$ and $\mu=1$.}
		\label{fig:SamplePaths}
	\end{center}
\end{figure}

The results in \Sref{sec:3} show convergence in the steady-state statistics of the $G/M/c$ to the steady-state statistics of the $G/M/\infty$. In this section, we study the convergence of the $G/M/c$ queueing system to the $G/M/\infty$ as $c\to\infty$ in a more general manner. A noticeable result is that, in the spatial configurations of the search process for which the $G/M/1$ queueing system blows up, a small number of additional servers shifts the behaviour of the system from uncontrolled growth to one of the $G/M/\infty$, recovering independent consumption. An example of this behaviour can be seen in \Fref{fig:SamplePaths}, where we simulated the sample paths for the $G/M/1$, $G/M/3$ and $G/M/\infty$ using the same arrival times and the same service times. The arrival times were simulated up to time $t=10,000$ and the queue length was simulated in the time interval $[0,8500]$. The simulation stores the arrival and departure times of each customer, the queue length at each of these times and the waiting time of each customer. The average waiting time in the simulations is 24.5411 for the $G/M/1$ and 0.0322 for the $G/M/3$. This significant reduction in the average waiting time, induced by the addition of two servers underlines the potential impact of a small number of additional servers. Moreover, the sample paths between the $G/M/3$ (\Fref{fig:SamplePaths}~(b)) and the $G/M/\infty$ (\Fref{fig:SamplePaths}~(c)) are almost indistinguishable, suggesting rapid convergence from the $G/M/c$ to a $G/M/\infty$, as the number of servers increases.

For the rest of this section, we assume that the spatial configuration of the search process will be such that for a given $L$ we fix the initial position to be $x_{0}^{*}$, the initial position that makes $\rho=1$ for the $G/M/1$ system; this is the largest initial position that makes the system blow up. First, with this spatial configuration, for each $c>1$, we have $\rho<1$ for the corresponding $G/M/c$. Therefore, by simply augmenting the number of servers in the system by one, we shift the behaviour from blow-up to relaxation to a steady state. In \Sref{sec:3}, we analysed how steady-state statistics behave for a growing number of servers. Here, we introduce a performance measure for the $G/M/c$ system compared with the $G/M/\infty$ system. 

As we saw in \Sref{sec:2}, the explicit expression for the stationary distribution leaves little space for analytical manipulation. The steady-state distribution of the $G/M/\infty$ queueing system is also complex to work with. These facts make working with the distance between said distributions a difficult task. However, we can shift our focus to the difference between the consumption protocols of the $G/M/c$ and the $G/M/\infty$ and their impact on the waiting time before service. The independent consumption protocol defining the $G/M/\infty$ system renders the waiting time before service of an arbitrary customer to zero.
In contrast, the sequential consumption protocol defining the $G/M/c$ allows the waiting time before service of an arbitrary customer to be greater than zero, as there is a limited number of servers. This characteristic difference suggests that the waiting time distribution should be used to define the desired performance measure. To do so, we examine the Wasserstein distance between the waiting time in the $G/M/c$ and the waiting time in the $G/M/\infty$.

\begin{figure}[h!]
	\begin{center}
		\includegraphics[width = 0.54\textwidth]{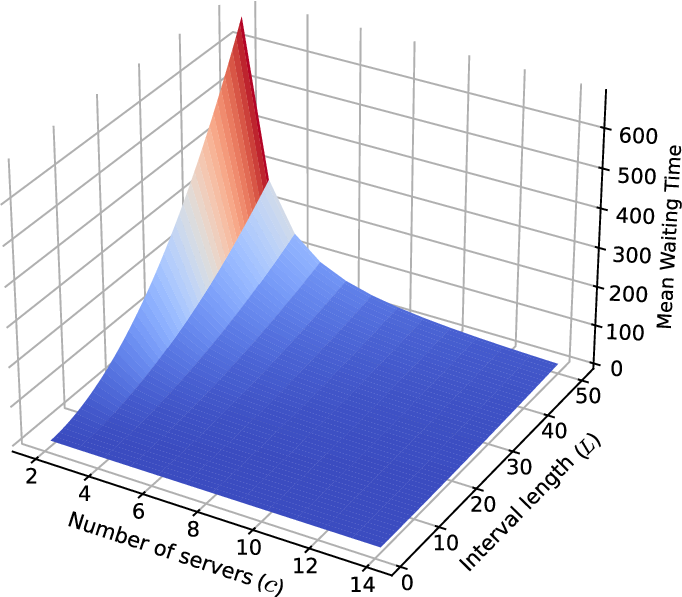}
		\caption{Systematic exploration of the Wasserstein distance between waiting time distributions for the $G/M/c$ and the $G/M/\infty$. For each value of $L$, we consider the critical initial position $x_{0}^{*}$ for which the $G/M/1$ has a traffic intensity of one. In every case, we fix $D=1$ and $\mu=1$.}
		\label{fig:3D}
	\end{center}
\end{figure}

The Wasserstein distance is generally defined for two probability measures $P,Q$ over $(X,\mathcal{B})$, where $X$ is a metric space with metric $d$ and $\mathcal{B}$ is the $\sigma$-algebra of Borel sets of $X$ in the following way:
\begin{equation*}
	R(P,Q) = \inf\{\mathbb{E}[d(\xi,\zeta)]\},
\end{equation*}
where the infimum is taken among all random variables $\xi$ and $\zeta$ with distributions $P$ and $Q$ respectively \cite{wasserstein1969markov, dobrushin_prescribing_1970}. This definition can be simplified when $X=\mathbb{R}$ with the usual Euclidean metric as follows
\begin{equation}\label{eq:WassersteinR1}
	R(P,Q) = \int_{-\infty}^{\infty}|F(x)-G(x)|d x,
\end{equation}
where $F$ and $G$ are the cumulative distribution functions of $P$ and $Q$ respectively \cite{vallender_calculation_1974}. We use this alternative definition to calculate the Wasserstein distance between the distribution of the waiting time until service in the $G/M/c$ queueing system, $F_c$, and the distribution of the waiting time until service in the $G/M/\infty$ queueing system, $F_\infty$. Given that the waiting time until service in the $G/M/\infty$ queueing system is always zero, we have
\begin{equation*}
	F_\infty(x) = \cases{
	0&for $x<0$,\\
	1&for $x\geq0$.}
\end{equation*}
However, for the $G/M/c$ queueing system using \eref{eq:WaitingTimeSurvival}, we have
\begin{equation*}
	F_c(x) = \cases{
	0&for $x<0$,\\
	1-\frac{A}{1-\sigma^{*}}e^{-(1-\sigma^*)c\mu x}&for $x\geq0$.
	}
\end{equation*}
Plugging in both distribution functions into \eref{eq:WassersteinR1} we have
\begin{equation}\label{eq:Wasserstein_Mean}
	\eqalign{
	R(F_\infty,F_c) & = \int_{0}^{\infty}\left|1-1-\frac{A}{1-\sigma^{*}}e^{-(1-\sigma^*)c\mu x}\right|d x\cr
	& = \int_{0}^{\infty}\frac{A}{1-\sigma^{*}}e^{-(1-\sigma^*)c\mu x}d x\cr
	& = \frac{A}{(1-\sigma^*)^2c\mu},
	}
\end{equation}
which is the mean waiting time before service of an arbitrary customer in the $G/M/c$ queueing system. Therefore, to compare the distribution of the performance of the $G/M/c$ with that of the $G/M/\infty$, it suffices to examine the mean waiting time before service. This simple measure of performance allows us to easily develop a systematic exploration of the performance of the $G/M/c$ compared with that of the $G/M/\infty$ for various numbers of servers and interval lengths, see \Fref{fig:3D}.

In \Fref{fig:3D}, we show a plot of the Wasserstein distance between the waiting time distributions for the $G/M/c$ and the $G/M/\infty$---equivalent to the mean waiting time before service in the $G/M/c$---as a function of the number of servers and the interval length. In this plot, for each interval length, $L$, we consider as initial position $x_{0}^{*}$, the value for which the $G/M/1$ traffic intensity is precisely one. The idea behind this particular selection of initial position is to analyse how adding servers takes the queueing system from one that is on the boundary of blowing up and whose mean waiting time before service diverges to one whose mean waiting time before service is finite. Moreover, we analyse the number of servers needed to have almost surely null waiting times before service. 

It is worth recalling from \Sref{sec:3} that $x_{0}^{*}=0$ as $L\to0$. Therefore, the scenarios explored in \Fref{fig:3D} are those of extreme spatial configurations for the search process, where $x_{0}/L\approx0$. For a fixed initial position, $x_{0}>x_{0}^{*}$, convergence to almost surely null waiting times before service occurs for a smaller number of servers. The fact that $x_{0}^{*}=0$ as $L\to0$ for each $c$ also explains the cusp observed in \Fref{fig:3D} at large $L$ and small $c$. Let us momentarily denote by $x_{0}^{*}(c,L)$ the critical initial position for the $G/M/c$ queue arising from a search process in an interval of length $L$. On one hand, as $x_{0}^{*}(c,L)=0$ as $L\to0$ for each $c$, $|x_{0}^{*}(1,L)-x_{0}^{*}(c,L)|=0$ as $L\to0$ for each $c>1$. On the other hand, from \eref{eq:ExistenceThreshold}, if $c_{1}>c_{2}$ then $x_{0}^{*}(c_{1},L)>x_{0}^{*}(c,L)$. The combination of these effects makes the critical initial position for which the $G/M/1$ traffic intensity is exactly one, $x_{0}^{*}$, approach from above the critical initial position for the $G/M/c$ at large $L$. This approach to the critical starting position occurs faster for systems with a small $c$, exponentially increasing their waiting times.
\section{Competition between two targets}\label{sec:5}
\begin{figure}[ht!]
	\begin{center}
		\includegraphics[width = 0.8\textwidth]{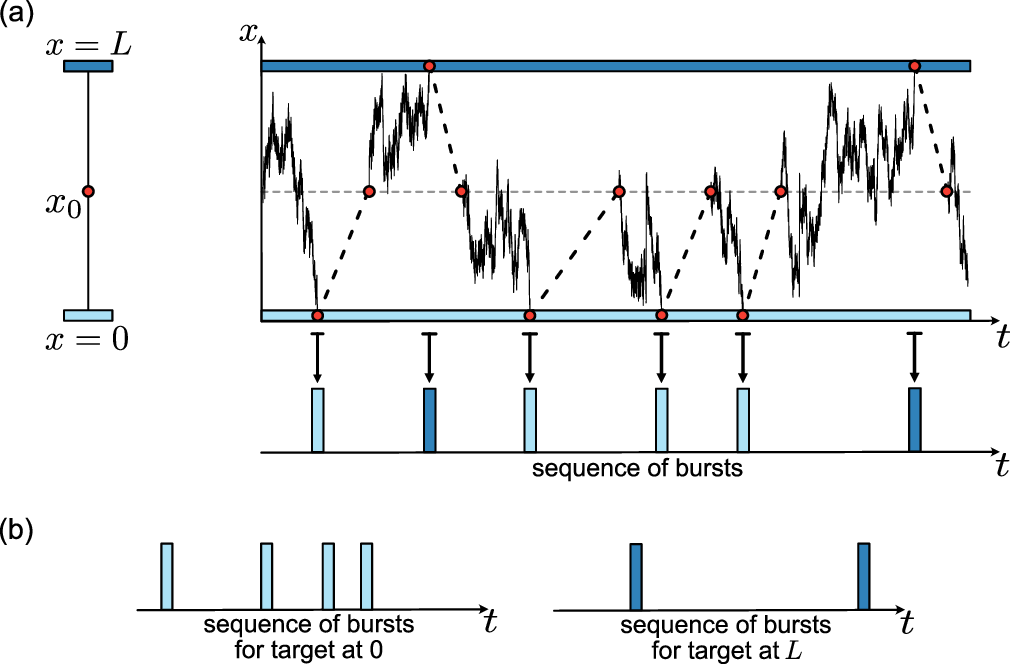}
		\caption{Accumulation of resources in two competing targets after several rounds of search-and-capture processes. (a) Multiple search-and-capture events for a particle diffusing inside the finite interval $[0,L]$ with two absorbing targets, one at the origin and one at $L$. Each time the particle finds a target, it delivers a resource (burst event) and returns to its initial position $x_0$, reloading and restarting the process (dash-dotted lines). (b) The total sequence of burst events now splits into two sequences, one for the target at the origin and one for the target at $L$. This splitting modifies the mean inter-arrival times for each of the queueing systems.}
		\label{fig:MultitargetmodelingScheme}
	\end{center}
\end{figure}

We now extend the analysis above to the case with two competing targets. The modelling rationale linking the accumulation of resources after several rounds of search-and-capture for each target remains the same as that for a single target. However, adding a target to the system at $x=L$ replaces the reflecting boundary with an additional absorbing boundary, see \Fref{fig:MultitargetmodelingScheme}. This introduces the splitting probabilities $\pi_{k}(x_{0})$ for the particle to be captured by the target at $k$, with $k\in\{0,L\}$ having started at $x_{0}$ and the conditional MFPT having started at $x_{0}$, $T_{k}(x_{0})$. These new quantities in the new search problem are given by
\begin{equation}\label{eq:SplittingProbabilities}
\pi_{0}(x_{0}) = 1-\frac{x_{0}}{L},\quad \pi_{L}(x_{0}) = \frac{x_{0}}{L},
\end{equation}
and
\begin{equation}\label{eq:ConditionalMFPT}
T_{0}(x_{0}) = \frac{2Lx_{0}-x_{0}^{2}}{6D},\quad T_{L}(x_{0}) = \frac{L^{2}-x_{0}^{2}}{6D}.
\end{equation}

The relationship between the conditional FPT densities, denoted by $f_{k}(t)$ with $k\in\{0,L\}$ representing the target, and the inter-arrival time density for the target $k$, denoted by $\mathcal{F}_{k}(t)$, is not as simple as in the case of a single target and requires additional analysis. This new relationship occurs because there can be an arbitrary number of arrivals to the competing target between each arrival to a particular target; see \Fref{fig:MultitargetmodelingScheme} for example. This relationship was explored in \cite{bressloff_modeling_2020, bressloff_search-and-capture_2019}.

We will derive an expression for the inter-arrival time density for target $k$, incorporating the effects of the loading and unloading times. To simplify the notation, for an index $k\in\{0,L\}$ representing one of the targets, we define $k^{c}$ as the complementary index representing the other target. Moreover, we drop the explicit dependence on $x_{0}$ from the notation of splitting probabilities. From the fact that between each arrival to a particular target there can be an arbitrary number of arrivals to the competing target, it follows that
\numparts
\begin{eqnarray}
	\fl\mathcal{F}_{k}(t) &= \pi_{k}\int_{0}^\infty f_{k}(t)\varphi(t-\tau)d\tau \nonumber\\ 
	 \fl&\quad+ \pi_{k}\pi_{k^{c}}\int_{0}^{t}\left(\int_{0}^{\infty} f_{k^{c}}(\gamma)\varphi(\tau-\gamma)d\gamma\right)\left(\int_{0}^{\infty} f_{k}(\gamma')\varphi(t-\tau-\gamma')d\gamma'\right)d\tau \label{eq:IATdensityA}\\
	 \fl&\quad+\dots \nonumber\\ 
	 \fl&= \pi_{k}(f_{k}*\varphi)(t) + \pi_{k}\pi_{k^{c}}\int_{0}^{t}(f_{k}*\varphi)(\tau)\cdot(f_{k^{c}}*\varphi)(t-\tau)d\tau + \dots\label{eq:IATdensityB}
\end{eqnarray}
\endnumparts
Laplace transforming this equation, we have
\begin{eqnarray}\label{eq:IATdensityLaplace}
	\fl\widetilde{\mathcal{F}}_{k}(s)  &= \pi_{k}\widetilde{f}_{k}(s)\widetilde{\varphi}(s) + \pi_{k}\pi_{k^{c}}[\widetilde{f}_{k}(s)\widetilde{\varphi}(s)][\widetilde{f}_{k^{c}}(s)
	\widetilde{\varphi}(s)] + \pi_{k}\pi_{k^{c}}^{2}[\widetilde{f}_{k}(s)\widetilde{\varphi}(s)][\widetilde{f}_{k^{c}}(s)\widetilde{\varphi}(s)]^{2}\nonumber\\
	\fl&\quad +\dots\\
	\fl&= \pi_{k}\widetilde{f}_{k}(s)\widetilde{\varphi}(s)\sum_{i=0}^{\infty}[\pi_{k^{c}}\widetilde{f}_{k^{c}}(s)\widetilde{\varphi}(s)]^{i}\nonumber.
\end{eqnarray}
Observe that $\widetilde{f}_{k^{c}}(s)\widetilde{\varphi}(s)$ is the Laplace transform of the probability density function of the conditional FPT to the target $k^{c}$. Therefore,
\begin{equation}\label{eq:LAP_Conditioned}
	\widetilde{f}_{k^{c}}(s)\widetilde{\varphi}(s) = \mathbb{E}[e^{-s\mathcal{T}_{k^{c}}}\mid \mathcal{T}_{k^{c}}<\infty],
\end{equation}
and, as $\mathcal{T}_{k^{c}}$ is a nonnegative random variable, we have $0<\widetilde{f}_{k^{c}}(s)\widetilde{\varphi}(s)<1$ for each $s>0$. Hence, \eref{eq:IATdensityLaplace} is a converging geometric series. Summing the resulting geometric series leads to the following expression
\begin{equation}\label{eq:IATdensity}
	\widetilde{\mathcal{F}}_{k}(s) = \frac{\pi_{k}\widetilde{f}_{k}(s)\widetilde{\varphi}(s)}{1-\pi_{k^{c}}\widetilde{f}_{k^{c}}(s)\widetilde{\varphi}(s)}.
\end{equation}

Denoting the inter-arrival times of the target at $k$ as $\{\Delta_{n}^{(k)}\}_{n\geq1}$, the mean inter-arrival time is given by
\begin{equation}\label{eq:MeanIATFormula}
	\mathbb{E}[\Delta_{n}^{(k)}] = -\frac{d}{ds}\widetilde{\mathcal{F}}_{k}(s)\Big|_{s=0} = T_{k}(x_{0}) + \tauc + \frac{\pi_{k^{c}}}{\pi_{k}}(T_{k^{c}}(x_{0})+\tauc).
\end{equation}
We now apply this equation to each of the targets to obtain the following mean inter-arrival times
\numparts
\begin{eqnarray}
	\mathbb{E}[\Delta_{n}^{(0)}] = \frac{Lx_{0}}{2D}+\tauc\left(1+\frac{x_{0}}{L-x_{0}}\right),\label{eq:MeanIATs0}\\
	\mathbb{E}[\Delta_{n}^{(L)}] = \frac{L(L-x_{0})}{2D}+\tauc\left(1+\frac{L-x_{0}}{x_{0}}\right).\label{eq:MeanIATsL}
\end{eqnarray}
\endnumparts
These two equations provide all the necessary information from the search process to replicate the blow-up analysis for a single target in \Sref{sec:2} in the case of two competing targets.

\subsection{Instantaneous refractory times}
First, we assume that once the searcher has found any of the two competing targets, it immediately unloads the resources, returns to the initial position, loads a new cargo and starts the new search-and-capture round; this is $\varphi(\hat{\tau}) = \delta(\hat{\tau})$ and $\tauc=0$. As the arrival rate for each target is the inverse of its corresponding mean inter-arrival times, \eref{eq:MeanIATs0} and \eref{eq:MeanIATsL}, we obtain the following expressions for the arrival rate of each target
\numparts
\begin{eqnarray}
	\lambda^{(0)} = \frac{2D}{Lx_{0}},\label{eq:ArrivalRate0}\\
	\lambda^{(L)} = \frac{2D}{L(L-x_{0})}.\label{eq:ArrivalRateL}
\end{eqnarray}
\endnumparts
Therefore, the condition for the existence of a stationary distribution of the $G/M/c$ system for each of the targets in terms of the traffic intensity \eref{eq:rho-leq1} becomes
\numparts
\begin{eqnarray}
	\rho^{(0)} = \frac{2D}{Lx_{0}c\mu}<1,\label{eq:TrafficIntensity0}\\
	\rho^{(L)} = \frac{2D}{L(L-x_{0})c\mu}<1.\label{eq:TrafficIntensityL}
\end{eqnarray}
\endnumparts
From these conditions, we independently obtain the threshold quantities for the interval length and initial position to ensure the existence of a steady-state distribution for each target. Equating the traffic intensities in \eref{eq:TrafficIntensity0} and \eref{eq:TrafficIntensityL} to one and solving each of the resulting expressions, we obtain
\numparts
\begin{eqnarray}
	x_0^{*(0)} = \frac{2D}{Lc\mu},\label{eq:x0s0}\\
	x_0^{*(L)} = L-\frac{2D}{Lc\mu}.\label{eq:x0sL}
\end{eqnarray}
\endnumparts

\begin{figure}[h!]
	\begin{center}
		\includegraphics[width = 0.65\textwidth]{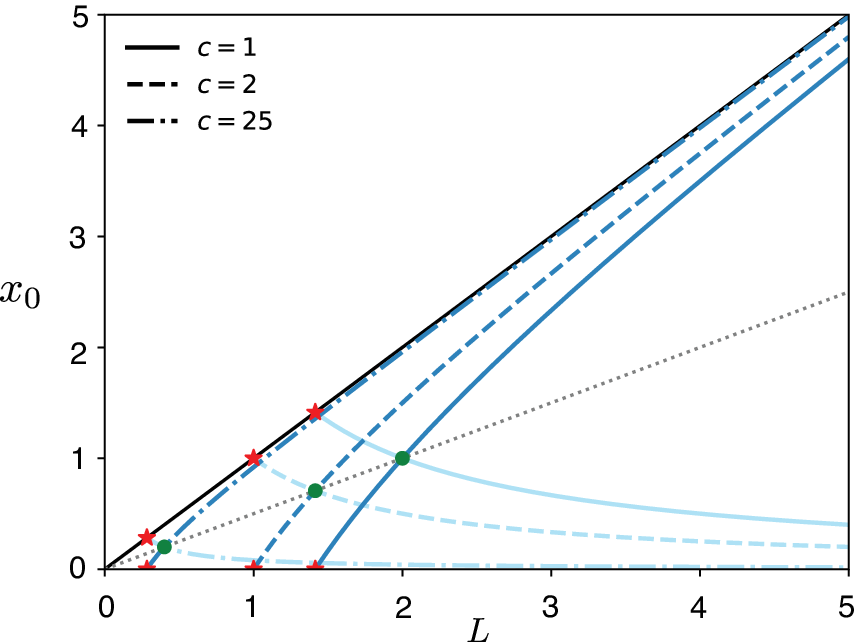}
		\caption{Critical spatial configurations for the existence of a steady-state distribution with two identical targets. The light curves represent the first target, located at $x=0$, and the dark curves represent the second target, located at the boundary $x=L$. The regions where the spatial configuration enables a steady state for both of the targets are determined by the intersection of the area above the light curve (existence of steady-state distribution for the first target) and the area below the dark curve (existence of steady-state distribution for the second target) for each specific number of servers. As in the case of one target, the red stars represent the maximum interval length for which a steady-state distribution for either of the two targets never exists. The green dots represent the unique combination $(\hat{L},\hat{x}_{0})$ for which both of the targets have a traffic intensity equal to one. This points are of the form $\hat{x}_{0} = \hat{L}/2$. In every case, we fix $D=1$ and $\mu=1$.}
		\label{fig:CriticalLinesMultiTarget}
	\end{center}
\end{figure}

In contrast to the scenario with a single target, both of these equations take real values for each $L>0$. However, there are values of $L\in(0,\infty)$ for which $x_0^{*(0)}>L$ and $x_0^{*(L)}<0$, leading to meaningless results as every starting position must be in $[0,L]$. It can be seen that $x_0^{*(0)}\leq L$ and $x_0^{*(L)}\geq0$ whenever
\begin{equation*}
	L\geq\sqrt{\frac{2D}{c\mu}},
\end{equation*}
and in those cases the critical initial positions $x_0^{*(0)}$ and $x_0^{*(L)}$ give meaningful information. Defining $L^{*}$ as the minimum length for which $x_0^{*(0)}\leq L$ and $x_0^{*(L)}\geq0$,
\begin{equation}\label{eq:CritialLTwoTargets}
	L^*: = \sqrt{\frac{2D}{c\mu}},
\end{equation}
it follows that for $L\geq L^{*}$, if $x_{0}\in(x_0^{*(0)},L]$ then $\rho^{(0)}\leq1$ and if $x_{0}\in[0,x_0^{*(L)})$ then $\rho^{(L)}\leq1$. Moreover, whenever $L<L^{*}$ we have $\rho^{(0)},\rho^{(L)} >1$ regardless of the starting position, $x_{0}$. Hence, to have a spatial setting that allows both systems to converge to a steady state, the interval length must fulfil $L>L^*$ and the initial position must fulfil $x_{0}^{*(0)}<x_{0}<x_{0}^{*(L)}$, see \Fref{fig:CriticalLinesMultiTarget}. Again,  $L^*$ and $x_{0}^{*(0)}$ vanish in the limit $c\to\infty$ and $x_{0}^{*(L)}$ converges to $L$ in the limit $c\to\infty$, recovering the fact that a system with an independent consumption protocol modelled by a $G/M/\infty$ has a steady-state distribution regardless of its spatial configuration.

In addition to these quantities, we can compute the unique interval length, $\hat{L}$, and initial position, $\hat{x}_{0}$, for which both queueing systems have a traffic intensity of exactly one. To do this, we first equate \eref{eq:x0s0} and \eref{eq:x0sL} and solve for $L$. This renders the following explicit expression
\begin{equation}\label{eq:Lhat}
	\hat{L} = 2\sqrt{\frac{D}{c\mu}}.
\end{equation}
Evaluating either the critical initial positions \eref{eq:x0s0} or \eref{eq:x0sL} in $\hat{L}$ yields the following value of $\hat{x}_{0}$
\begin{equation}\label{eq:Xhat}
	\hat{x}_{0} = \sqrt{\frac{D}{c\mu}} = \frac{\hat{L}}{2}.
\end{equation}
This equation shows that the only way both queueing systems can have a unitary traffic intensity is for the initial position of the search process to be in the midpoint of the interval with the appropriate length $\hat{L}$. Furthermore, it highlights the fact that the behaviour of both targets, depending on the initial position, is symmetric with respect to the midpoint of the interval, see \Fref{fig:CriticalLinesMultiTarget}. We omit the analysis of the steady-state statistics in this scenario, as it is extremely similar to the analysis made for the single target scenario. 

Now, we focus exclusively on the target at the origin and consider the effects of adding a target at the other end to its traffic intensity. Intuition would say that adding a target would introduce competition, and this competition takes away resources from the original target, decreasing traffic intensity. However, if the searcher has instantaneous unloading and loading times, adding a target increases the traffic intensity at the original target. (For ease of reference, we include the time to return to $x_0$ in the loading time.) This situation translates into the following inequality between the traffic intensity of the target at the origin in the single-target scenario \eref{eq:ExistenceCondition} and the traffic intensity of the target at the origin in the two-target scenario \eref{eq:TrafficIntensity0}
\begin{equation}\label{eq:TrafficIntensitiesCompetence}
	\rho = \frac{2D}{c\mu(2L-x_{0})x_{0}}\leq\frac{2D}{c\mu Lx_{0}}=\rho^{(0)},\qquad L>0,\,\,x_{0}\in[0,L].
\end{equation}
This reflects the fact that competition from the second target is counteracted by the instantaneous return to $x_0$ following absorption at $x=L$, compared to the slow return from a reflecting boundary at $x=L$. That is, when the searcher has instantaneous refractory times, adding one target improves the delivery rate of the searcher to the original target at $x=0$. We briefly explore the effects of non-instantaneous unloading and loading times on improving the delivery rate and traffic intensity for the target at the origin.

\subsection{Effects of nonzero refractory times}
\begin{figure}[ht!]
	\begin{center}
		\includegraphics[width = 0.65\textwidth]{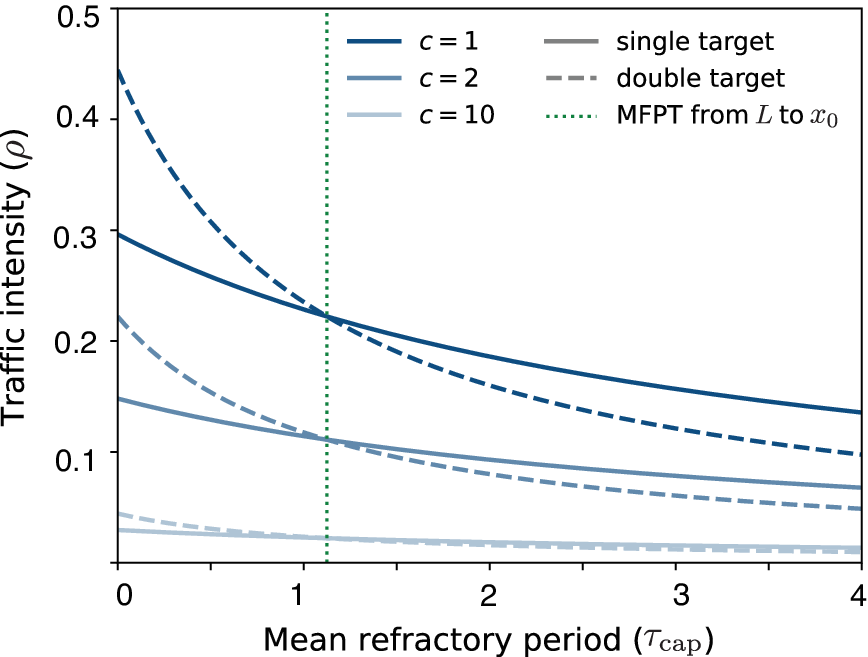}
		\caption{Effects of the mean refractory time $(\tauc)$ on the traffic intensity of the queueing system associated with the target at $x=0$ in the scenarios of a single target and two targets. Whenever the mean refractory time is less than the MFPT from $L$ to $x_{0}$ (dotted line), the traffic intensity of the target at $x=0$ in the scenario of two competing targets (dashed lines) is greater than the traffic intensity of the target at $x=0$ in the scenario of a single target (solid lines). In every case, we fix $D=1$, $\mu=1$, $L=3$ and $x_{0}=1.5$.}
		\label{fig:CriticalLinesRef}
	\end{center}
\end{figure}

We now extend the analysis to include nonzero refractory times $\hat{\tau}$ with density $\varphi(\hat{\tau})$ and mean $\tauc$. This modifies the inter-arrival times of the queueing process associated with the target at the origin. We equate the right-hand side of \eref{eq:MeanIATs0} to $T(x_{0})+\tauc$, where $T(x_0)$ is given by equation (\ref{eq:MFPT}). Solving for $\tauc$, we obtain the following mean refractory time for which the inter-arrival times of the queueing process associated with the target at the origin are the same in the single-target and two-target scenarios:
\begin{equation}\label{eq:TaucThreshold}
	\tauc^{*} = \frac{(L-x_{0})^{2}}{2D}.
\end{equation}
The right-hand side of this equation precisely coincides with the MFPT from the reflecting boundary at $L$ to the initial position $x_{0}$ in the single-target scenario; see \Fref{fig:CriticalLinesRef}. In other words, if the mean time for the searcher to return from the reflecting boundary to the initial position in the single-target scenario is equal to the mean refractory time in the two-target scenario, the mean inter-arrival times are the same in both cases.

In \Fref{fig:CriticalLinesRef}, we show plots for the traffic intensity of the queueing system associated with the target at $x=0$ in the scenarios of a single target and two targets as a function of the mean refractory time, $\tauc$. This plot shows that when the mean refractory time, $\tauc$, is less than the MFPT from the reflecting boundary at $L$ to the initial position $x_{0}$ in the single-target scenario, the traffic intensity of the queueing system associated with the target at $x=0$ is greater in the two-target scenario. In these cases, competition between targets becomes an advantage for the original target. However, when the mean refractory time, $\tauc$, is greater than the MFPT from the reflecting boundary at $L$ to the initial position $x_{0}$ in the single-target scenario, the traffic intensity of the queueing system associated with the target at $x=0$ is greater in the single-target scenario. Competition reduces the delivery rate of the original target.

\section{The effects of multiple searchers}\label{sec:6}
We end our analysis by considering multiple searchers and a single target. We derive analytical results for the general $G/M/c$ model and show the numerical results for $G/M/1$, compared with $G/M/\infty$. However, extending the numerical results to scenarios with two targets and multiple servers is straightforward. We assume that $M\in\mathbb{N}$ independent searchers deliver resources to a single target under the conditions specified in \Sref{sec:2}. The searchers are assumed to have the same initial position, $x_{0}$, and diffusivity $D=1$.

We define the sequence of burst events associated with the target by combining all the sequences of burst events produced by each searcher. This new sequence of burst events modifies the inter-arrival time density, defining the queueing system associated with the target. The superposition of renewal processes must be studied to determine the new inter-arrival time density. Generally, the superposition of renewal processes does not necessarily define a new renewal process, which makes the analysis of the superposition of renewal processes somewhat complex and has motivated a significant amount of work on approximations and their direct relation to queueing theory \cite{albin_poisson_1982, albin_approximating_1984, whitt_approximating_1982, lam_superposition_1991}. Fortunately, we are only interested in determining the inter-arrival time density generated by the superposition of $M$ identical searchers, which can be achieved using classical renewal theory \cite{mitov_renewal_2014}.

We begin by considering a family of inter-arrival time sequences
\begin{equation}\label{eq:IAT_Family}
	\mathcal{I} := \bigcup_{i\geq1}\{\Delta_{n}^{(i)}\}_{n\ge0},
\end{equation}
where $F^{(i)}_{\Delta}(t)$ is the distribution of the $i$th inter-arrival time sequence. Each inter-arrival time sequence independently defines a renewal process $\{S_{n}^{(i)}\}_{n\geq0}$, where $S_{0}^{(i)}=0$ and $S_{n}^{(i)}:=\Delta_{1}^{(i)}+\dots+\Delta_{n}^{(i)}$. We consider $\{\mathcal{S}_{n}^{(M)}\}_{n\geq0}$ as the sequence of renewal epochs obtained by considering the union of the first $M$ renewal processes $\{S_{n}^{(i)}\}_{n\geq0}$ and arranging them in increasing order. This represents the renewal epochs, or customer arrivals at a target, coming from any of $M$ different sources or searchers. For further details, see \ref{ap:Renewal}. Denoting by $\gamma$ the mean inter-arrival time of each burst sequence, the resulting combined sequence of bursts $\{\mathcal{S}_{n}^{(M)}\}_{n\geq0}$ has a mean inter-arrival time $\gamma M^{-1}$. This expression for the mean inter-arrival time implies that traffic intensity increases linearly with the number of searchers
\begin{equation}\label{eq:TrafficIntensityMulti}
	\rho^{(M)} = \frac{M}{c\mu T(x_{0})}. 
\end{equation}
Given this linear dependence of the traffic intensity, it is straightforward to modify the expression for the critical initial position $x_{0}$ of a single searcher \eref{eq:ExistenceThreshold} to obtain the following expression for the critical initial position of $M$ searchers
\begin{equation}\label{eq:ExistenceThresholdMultiSearcher}
	x_0^*(M) = L-\sqrt{L^2-\frac{2DM}{c\mu}}.
\end{equation}

Analogous to the scenario with a single target and a single searcher, we obtain the maximal interval length for which the system with $M$ searchers does not converge to a steady state, regardless of the searchers' initial position, as the value of $L$ for which the discriminant of the quadratic equation resulting from equating \eref{eq:TrafficIntensityMulti}  to unity is exactly zero:
\begin{equation}\label{eq:CritialLMultiSearcher}
L^*(M) := \sqrt{\frac{2DM}{c\mu}}.
\end{equation}
Again, note that if $L<L^*(M)$ then the quadratic equation resulting from equating \eref{eq:TrafficIntensityMulti} to unity has no real roots. Moreover, whenever $L<L^*(M)$ we have $\rho^{(M)}>1$ regardless of the starting position $x_{0}$. Finally, in the limiting case $L=L^*(M)$, any configuration makes the system blow up since $x_{0}^{*}(M)=L^{*}(M)$.

\begin{figure}[h!]
	\begin{center}
		\includegraphics[width = 0.65\textwidth]{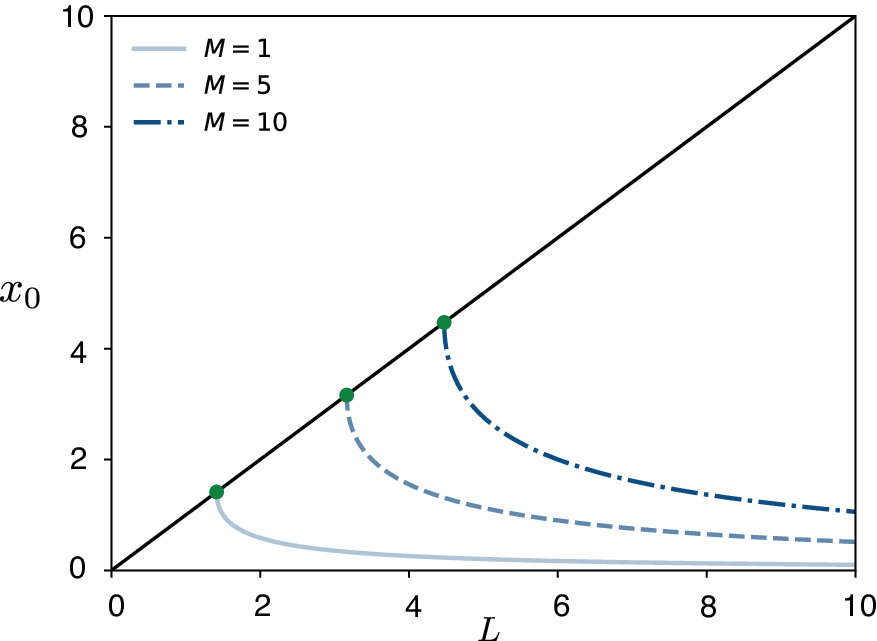}
		\caption{Critical spatial configurations for the existence of a steady-state distribution for the $G/M/1$ with multiple searchers. Each curve represents the threshold spatial configurations for the search-and-capture process that make the $G/M/1$ queueing system for the accumulation of resources blow up. These curves correspond to \eref{eq:ExistenceThresholdMultiSearcher} with 1, 5 and 10 searchers in a $G/M/1$ queueing system. Any pair of values $(L,x_{0})$ below the diagonal and above the curve corresponding to a $M$ defines a spatial configuration that allows the queueing system with said number of searchers to relax to a steady state. The green points represent the maximum interval length for which a steady-state distribution never exists in each case. In every case, we fix $D=1$ and $\mu=1$.}
		\label{fig:CriticalLinesMultiSearcher}
	\end{center}
\end{figure}

In \Fref{fig:CriticalLinesMultiSearcher}, we see how increasing the number of searchers has the opposite effect compared to increasing the number of servers, making the critical initial position larger and reducing the space of possible spatial configurations that ensure convergence to a steady state. Now, $L^{*}$ and $x_{0}^{*}$ explicitly depend on the number of servers, $c$, and the number of searchers, $M$, but we drop the explicit dependence from the notation for the sake of simplicity. For each fixed $M\in\mathbb{N}$, the limiting behaviour of these quantities is preserved as both $L^*$ and $x_0^*$ vanish in the limit $c\to\infty$ and $x_{0}^{*}$ vanishes in the limit $L\to\infty$.

Having determined the spatial configurations that allow the system with $M$ searchers to converge to a steady state, we now analyse the steady-state statistics for the queue length. Recall that the steady-state statistics of the $G/M/c$ and $G/M/1$ depend on the generalised traffic intensity, determined by the implicit equation \eref{eq:sigma_star} in terms of the Laplace transform of the inter-arrival time density. Therefore, to make sense of the expressions in \eref{eq:GM1_Moments} and \eref{eq:GMc_Moments}, we first need a method to compute the Laplace transform of the inter-arrival time density in the scenario with $M$ searchers, denoted by $\widetilde{\mathcal{F}}_{M}$. First, recall that 
\begin{equation*}
\mathcal{F}_{M}(t)=-\frac{d}{dt}Q_{M}(t),
\end{equation*}
where $Q_{M}$ denotes the survival function of the inter-arrival times for the process with $M$ searchers, and $\widetilde{\mathcal{F}}_{M}(s)=1-s\widetilde{Q}_{M}(s)$. Given these identities, we focus on computing $\widetilde{Q}_{M}(s)$.

As mentioned before, we can compute the survival function of the inter-arrival times for the superposition process using classical renewal processes theory. For further details, refer to \ref{ap:Renewal}. We have
\begin{equation}\label{eq:SurvivalSuperpositionM}
	Q_{M}(t) = Q_{\Delta}(t)\left(\frac{1}{\gamma}\int_{t}^{\infty}Q_{\Delta}(u)du\right)^{M-1},
\end{equation}
where $Q_{\Delta}$ is the survival function of the inter-arrival times for the process with a single searcher. We can easily compute the Laplace transform of $Q_{\Delta}$ by using the results presented in \Sref{sec:2}. However, there is no explicit expression for the function $Q_{\Delta}$ itself. Therefore, we resort to numerical computations to compute $\widetilde{Q}_{M}(s)$. These computations are performed numerically calculating the inverse Laplace transform of $\widetilde{Q}_{M}(s)$, then numerically integrating $Q_{M}(s)$ from $t$ to $\infty$, and finally numerically computing the Laplace transform on the right-hand side of \eref{eq:SurvivalSuperpositionM}. This allows us to numerically compute $\widetilde{\mathcal{F}}_{M}$ and determine the generalised traffic intensity $\sigma^{*}$. Given the generalised traffic intensity, we can use analytic expressions for the mean and variance of the queue length at steady state in the $G/M/1$ and the mean of the queue length at steady state in the $G/M/c$.

From the results in \cite{bressloff_queueing_2020}, we can directly obtain the steady-state statistics of the $G/M/\infty$ queue with $M$ searchers as
\begin{equation}\label{eq:MeanGMInfty}
	\mathbb{E}[Q_{\infty}^{(\infty)}] = \frac{\lambda}{\mu} = \frac{M}{\mu T(x_{0})},
\end{equation}
and
\begin{equation}\label{eq:VarGMInfty}
	\fl\mathrm{Var}[Q_{\infty}^{(\infty)}] = \frac{\lambda}{\mu}\left[\frac{\widetilde{\mathcal{F}}_{M}(\mu)}{1-\widetilde{\mathcal{F}}_{M}(\mu)}+1-\frac{\lambda}{\mu}\right]=\frac{M}{\mu T(x_{0})}\left[\frac{\widetilde{\mathcal{F}}_{M}(\mu)}{1-\widetilde{\mathcal{F}}_{M}(\mu)}+1-\frac{M}{\mu T(x_{0})}\right].
\end{equation}
Therefore, we can compute the mean directly from \eref{eq:MeanGMInfty} and, using the numerical computations of $\widetilde{\mathcal{F}}_{M}$, we can compute the variance from \eref{eq:VarGMInfty}. Finally, we compare the behaviour of the steady-state statistics for $G/M/1$ and the $G/M/\infty$ as a function of the number of searchers.

\begin{figure}[h!]
	\begin{center}
		\includegraphics[width = \textwidth]{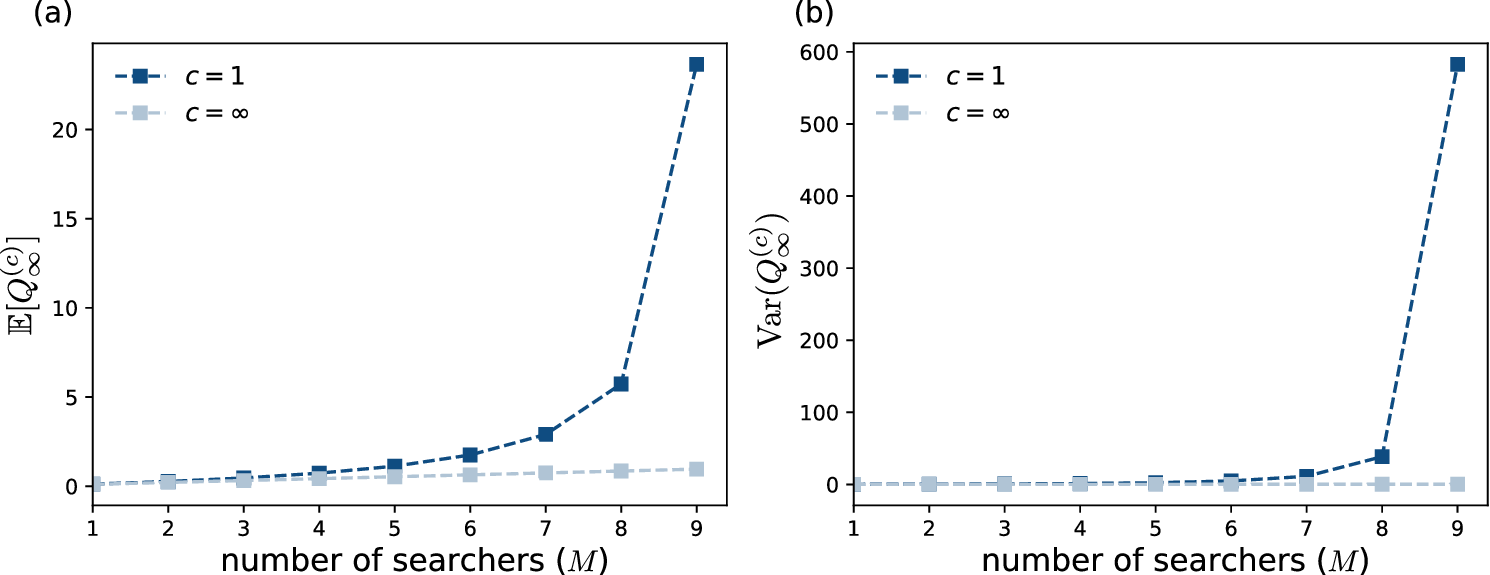}
		\caption{Comparison of mean and variance of the queue length at steady state between the $G/M/1$ and the $G/M/\infty$ for different numbers of searchers. (a) Plots of the steady-state mean, $\mathbb{E}[Q_\infty^{(c)}]$, as a function of the number of searchers. (b) Plots of the steady-state variance, Var$(Q_\infty^{(c)})$, as a function of the number of searchers. The number of searchers is up to nine since the traffic intensity for more searchers is greater than one, and the $G/M/1$ system blows up. In every case we fix $L=5$, $x_{0}=2.5$, $D=1$ and $\mu=1$.}
		\label{fig:MeanVarMultiSearcher}
	\end{center}
\end{figure}

\Fref{fig:MeanVarMultiSearcher}(a), shows plots of the mean number of resources in the $G/M/1$ and $G/M/\infty$ systems at steady state as a function of the number of searchers. The steady-state mean for the $G/M/1$ is computed using the explicit formulas of steady-state statistics \eref{eq:GM1_Moments} with generalised traffic intensity, $\sigma^{*}$, determined using the numerical procedure above. The steady-state mean of the $G/M/1$ grows exponentially as the number of searchers increases, while the steady-state mean of the $G/M/\infty$ depends linearly on the number of searchers. Moreover, the exponential growth of the steady-state mean of the $G/M/1$ occurs as the traffic intensity linearly grows to one from below. Once the number of searchers makes the traffic intensity greater than one, the $G/M/1$ blows up. However, the $G/M/\infty$ never blows up, and its mean keeps on growing linearly with the number of searchers. Corresponding plots for the variance are shown in \Fref{fig:MeanVarMultiSearcher}(b), and we observe a similar behaviour to the one just described for the mean.

\section{Discussion}
In this study, we extended the general framework for studying the accumulation of resources in a target under multiple rounds of search-and-capture events combined with resource consumption. Rather than modelling independent consumption in terms of a $G/M/\infty$ queue, as developed in \cite{bressloff_queueing_2020}, we considered sequential consumption based on a $G/M/c$ queue with $G/M/1$ as a special case. Two major differences between the $G/M/\infty$ and a $G/M/c$ queues are that in the latter case there are (i) constraints on the existence of a steady-state distribution and (ii) nonzero random waiting times. One of our main findings is the identification of critical regions for the interval length and initial position of the search process, defining the spatial conditions that ensure convergence to a steady-state distribution for the $G/M/c$ model. Under these conditions, we could find explicit expressions for the mean and variance of the $G/M/1$ queue, which only depend on the MFPT, consumption rate and generalised traffic intensity. These are given by \eref{eq:GM1_Moments} and \eref{eq:GM1_Var}, respectively. Using the Wasserstein distance, we also proved that the mean waiting time before service in the $G/M/c$ model can be used as a performance measure compared with the previously defined $G/M/\infty$ queue model. With this, we observed how a small number of additional servers causes the waiting times to vanish and almost surely allows the $G/M/c$ to recover independent consumption.

We also modified the basic $G/M/c$ queue model with a single target to consider two competing targets and multiple searchers. Regarding the two competing targets, we observed that whenever the mean refractory time is less than the MFPT from the reflecting boundary at $L$ to the initial position, $x_{0}$, the competition introduced by the second target benefits the original target at the origin. Finally, in the case of multiple searchers, we were able to numerically compute the generalised traffic intensity and utilise the explicit expressions for the mean and variance of the $G/M/1$ and the $G/M/\infty$. These expressions demonstrated an exponential growth of the mean queue length and the queue length variance of the $G/M/1$ at steady state as the number of searchers increases and the traffic approaches one. This highlights a relevant behavioural distinction between the $G/M/1$ and the $G/M/\infty$, as the mean queue length and the queue length variance of the $G/M/\infty$ at steady state increases linearly as the number of searchers increases and the traffic intensity approaches one.

There are several directions for future research in this area. First, the analysis can be extended to higher-dimensional search processes including concentric spheres and wedge domains. In particular, it would be interesting to determine how the critical regions for the existence of steady-state queue length distributions depend on the geometry of the search domain and the locations of one or more targets. Another modification of the search process is to include the effects of stochastic resetting, whereby the position of the searching particle is reset to a fixed position, $x_{R}$, with rate $r$ \cite{evans_stochastic_2020}. Stochastic resetting has been observed to improve the MFPT of a diffusive search process \cite{evans_diffusion_2011, evans_diffusion_2011-1, evans_optimal_2013}, with an optimal resetting rate $r^{*}$ that minimises the MFPT \cite{evans_stochastic_2020, evans_optimal_2013}. The first direction is to incorporate stochastic resetting in the interval, as done before in \cite{pal_first_2019}, and analyse the effects of the resetting rate in the critical regions defining the blow-up configurations. Stochastic resetting renders the MFPT in unbounded domains of arbitrary dimension finite \cite{evans_diffusion_2014}. By incorporating stochastic resetting, we can study the accumulation of resources in the $G/M/c$ queue emerging from freely diffusing particles in unbounded domains of higher dimension. This introduces an additional source of refractory times that complements those associated with the unloading and loading events.

An important property of a queue that we did not consider in this study is the existence of idle periods, defined as the periods in which the queue is empty. Several authors have studied the distribution of idle periods \cite{lopker_idle_2010, nazarathy_busy_2022, rosenlund_queue_1978, wolff_idle_2003 }. This is particularly important in scenarios of heavy traffic when $\rho\to1$ from below \cite{kingman_single_1961, kingman_queues_1962, whitt_approximations_1993, whitt_heavy-traffic_2005, gaunt_steins_2020}. In future work, we would like to determine the conditions for the search process that allow the target to relax to a steady state while maintaining a constant resource supply. A more recent perspective has been to incorporate resetting into the queue length \cite{roy_queues_2024} and service times \cite{bonomo_mitigating_2022, bonomo_queues_2025}. These modifications allow for new nontrivial dynamics in the queue length. It would be interesting to explore the effects of stochastic resetting on idle periods under scenarios of heavy traffic and scenarios of traffic intensity greater than one. Yet another future direction would be to apply the model to axonal transport and resource accumulation in synapses. For example, are there analogs of critical regions and blow up in the break down of synaptic function? It is also likely that the activity-driven consumption of  synaptic vesicles is non-Markovian, requiring the analysis of a $G/G/c$ queue.

\ack 
Jos\'{e} Giral-Barajas was supported by a Roth PhD scholarship funded by the Department of Mathematics at Imperial College London.


\appendix
\section{$G/M/c$ steady-state distribution}\label{ap:GMC}
\subsection{Embedded Markov chain}
Consider the $G/M/c$ queue consisting of $c$ servers $(1\leq c<\infty)$, in which individual customers arrive according to a general distribution $F_{\Delta}(t)$ and the time to service a customer is exponentially distributed with intensity $\mu$. Let $Q^{(c)}(t)$ be the number of customers in the system at time $t$, $\tau_{n}$ be the time of arrival of the $n$th customer, and let $Q^{(c)}_{\tau_{n}}=Q^{(c)}(\tau_{n}^-)$ be the number of customers waiting in the line ahead of the $n$th customer just prior to the time of their arrival. Here $\tau_n^-=\lim_{\epsilon \rightarrow 0^+}(\tau_n-\epsilon)$. We then obtain the following iterative equation
\begin{equation}\label{eq:GMC_Markov_AP}
	Q^{(c)}_{\tau_{n+1}} = Q^{(c)}_{\tau_{n}}+1-V^{(c)}_{n},
\end{equation}
where $V^{(c)}_{n}$ is the number of customers served between $[\tau_{n},\tau_{n+1})$. Note that $V^{(c)}_{n}$ depends on $Q^{(c)}_{\tau_{n}}$, since no more than $Q^{(c)}_{\tau_{n}}+1$ individuals can depart during $[\tau_{n},\tau_{n+1})$. However, since $Q^{(c)}_{\tau_{n}}$, $V^{(c)}_{n}$ are independent of $Q^{(c)}_{\tau_{n-1}},\dots,Q^{(c)}_{\tau_{0}}$, it follows that \eref{eq:GMC_Markov_AP} represents a discrete-time Markov chain embedded in the queue length process. Note that this discrete-time Markov chain jumps every time a new customer arrives at the queueing system, and it takes the value of the number of customers the arriving customer finds ahead of them in the system. Therefore, its steady-state distribution is often called the \textit{arriving customer distribution}. We now follow the results in chapter 5.14 from \cite{cooper_introduction_1981} to compute the transition probabilities of the embedded Markov chain
\begin{equation}\label{eq:GMC_probabilities_AP}
	\pi^{(c)}_{ij} = \mathbb{P}[Q^{(c)}_{\tau_{n+1}} = j\mid Q^{(c)}_{\tau_{n}}=i].
\end{equation}
\subsection{Transition probabilities}
First, observe that any arriving customer can find at most one more customer present in the system than as found by the preceding arriving customer. This corresponds to the extreme case where there are two consecutive arrivals without any departures in between. Hence
\begin{equation}\label{eq:GMC_probabilities_C1_AP}
	\pi^{(c)}_{ij} = 0,\qquad(j>i+1).
\end{equation}
Now, if the $(n+1)$th arriving customer finds all of the $c$ servers busy then the $n$th arriving customer could have found at most one idle server and all of the $c$ servers must have been continuously busy during the inter-arrival time $\Delta_{n+1}=\tau_{n+1}-\tau_{n}$. Since the service times are assumed to be exponentially distributed with intensity $\mu$, it follows that the probability that there are $k$ departures in an inter-arrival time of length $t$, given that there are $i$ customers is
\begin{equation}\label{eq:GMC_probabilities_C2_V_AP}
	\fl\mathbb{P}[V^{(c)}_{n}=k\mid Q^{(c)}_{\tau_{n}}=i,\Delta_{n+1}=t]=\frac{(c\mu t)^{k}}{k!}e^{-c\mu t},\,\,\,(i\geq c-1,\,0\leq k\leq i).
\end{equation}
Using the iterative equation for the evolution of the embedded Markov chain, \eref{eq:GMC_Markov_AP}, we can rewrite the probability above as
\begin{eqnarray}\label{eq:GMC_probabilities_C2_Q_AP}
	\fl\mathbb{P}[Q^{(c)}_{\tau_{n+1}}=j\mid Q^{(c)}_{\tau_{n}}=i,\Delta_{n+1}=t]=\frac{(c\mu t)^{i+1-j}}{(i+1-j)!}e^{-c\mu t},\\
\qquad\qquad\qquad\qquad\qquad\qquad\qquad\qquad(i\geq c-1,\,j\geq c,\,j\leq i+1),\nonumber
\end{eqnarray}
and since the inter-arrival times have distribution $F_{\Delta}(t)$ we finally obtain
\begin{equation}\label{eq:GMC_probabilities_C2_AP}
	\fl\pi^{(c)}_{ij}=\int_{0}^{\infty}\frac{(c\mu t)^{i+1-j}}{(i+1-j)!}e^{-c\mu t}dF_{\Delta}(t),\qquad(i\geq c-1,\,j\geq c,\,j\leq i+1).
\end{equation}

To determine the conditions for the existence of the steady-state distribution we only need the transition probabilities determined in \eref{eq:GMC_probabilities_C1_AP} and \eref{eq:GMC_probabilities_C2_AP}. However, for the sake of completion we state the remaining transition probabilities
\begin{equation}\label{eq:GMC_probabilities_C3_AP}
	\fl\pi^{(c)}_{ij}=\int_{0}^{\infty}{i+1\choose j}e^{-j\mu t}(1-e^{\mu t})^{i+1-j}dF_{\Delta}(t),\quad(i\leq c-1,\,j\leq i+1),
\end{equation}
and
\begin{eqnarray}\label{eq:GMC_probabilities_C4_AP}
	\fl\pi^{(c)}_{ij}=\int_{0}^{\infty}\int_{0}^{t}{c\choose j}e^{-j\mu(t-y)}(1-e^{-\mu(t-y)})^{c-j}\frac{(c\mu t)^{i-c}}{(i-c)!}e^{-c\mu y}c\mu\,dF_{\Delta}(t),\\
\qquad\qquad\qquad\qquad\qquad\qquad\qquad\qquad\qquad(i\geq c,\,j<c,\,j\leq i+1).\nonumber
\end{eqnarray}
\subsection{Steady-state arriving customer distribution}\label{subap:A3}
Now that we have the transition probabilities of the embedded Markov chain, we can determine the stationary steady-state arriving customer distribution (assuming it exists). Suppose there exists a unique steady-state arriving customer distribution
\begin{equation}\label{eq:EmMC_limit_AP}
	\Pi^{(c)}_j = \lim_{n\to\infty} \mathbb{P}[Q^{(c)}_{\tau_{n}} = j],
\end{equation}
which satisfies the steady-state distribution equations
\begin{equation}\label{eq:EmMC_EqCond_AP}
	\Pi^{(c)}_{j}=\sum_{i=0}^{\infty}\pi^{(c)}_{ij}\Pi^{(c)}_{i},\qquad\sum_{j=0}^{\infty}\Pi^{(c)}_{j}=1.
\end{equation}
Let us consider the first steady-state distribution equation for the values with index $j\geq c$. Substituting the explicit expressions for $\pi^{(c)}_{ij}$ given in \eref{eq:GMC_probabilities_C1_AP} and \eref{eq:GMC_probabilities_C2_AP} yields
\begin{eqnarray}\label{eq:EmMC_EqCond2_AP}
	\Pi^{(c)}_{j} & =\sum_{i=j-1}^{\infty}\Pi^{(c)}_{i}\int_{0}^{\infty}\frac{(c\mu t)^{i+1-j}}{(i+1-j)!}e^{-c\mu t}dF_{\Delta}(t),\nonumber\\
			  & =\sum_{i=0}^{\infty}\int_{0}^{\infty}\Pi^{(c)}_{i+1-j}\frac{(c\mu t)^{i}}{i!}e^{-c\mu t}dF_{\Delta}(t),\quad(j\geq c).
\end{eqnarray}
As a final step, consider the trial solution $\Pi^{(c)}_{j}=A(\sigma^{*})^{j-c}$ for $j\geq c-1$ and some $\sigma^{*}\in(0,1)$ that shall be determined. Substituting this trial solution into \eref{eq:EmMC_EqCond2_AP} we obtain
\begin{equation}\label{eq:EmMC_SSCond_AP}
	\sigma^{*}=\int_{0}^{\infty}e^{-(1-\sigma^{*})\mu t}dF_{\Delta}(t)=\widetilde{\mathcal{F}}(c\mu(1-\sigma^*)).
\end{equation}
Therefore, $\sigma^*$ is determined by the implicit equation
\begin{equation}\label{eq:sigma_star_AP}
	\sigma^* = \widetilde{\mathcal{F}}(c\mu(1-\sigma^*)),
\end{equation}
which we have shown to have a unique solution $\sigma^{*}\in(0,1)$ if and only if $\rho = \lambda/(c\mu)<1$. Observe that in the particular case of the $G/M/1$, when $c=1$, this determines the whole steady-state arriving customer distribution, given by
\begin{equation}\label{eq:Pi_GM1_AP_1}
	\Pi^{(1)}_{j}=A(\sigma^{*})^{j-1},\qquad(j\geq 0).
\end{equation}

The normalising constant $A$ can be simply determined using the second steady-state distribution equation \eref{eq:EmMC_EqCond_AP} such that $A=\sigma^{*}(1-\sigma^{*})$. Therefore, the steady-state arriving customer distribution for the $G/M/1$ is a geometric distribution with parameter $1-\sigma^{*}$
\begin{equation}\label{eq:Pi_GM1_AP}
	\Pi^{(1)}_{j}=(1-\sigma^{*})(\sigma^{*})^{j},\qquad(j\geq 0).
\end{equation}
Whenever $c\geq2$ the procedure above only defines $\Pi^{(c)}_{j}$ for $j\geq c-1$ and further steps need to be taken to determine the remaining probabilities $\Pi^{(c)}_{0},\dots\Pi^{(c)}_{c-2}$ and the value of the normalising constant $A$. In the rest of this section, we will outline these steps and compute a general expression for the remaining probabilities and the normalising constant, following the results presented in chapter 2 from \cite{takacs_introduction_1962}. 

First, define the following generating function
\begin{equation}\label{eq:Gen_Fun_AP}
	U^{(c)}(z)=\sum_{j=0}^{c-1}\Pi^{(c)}_{j}z^{j}.
\end{equation}
Using the first steady-state distribution equation \eref{eq:EmMC_EqCond_AP}, it can be shown that
\begin{eqnarray}\label{eq:Gen_Fun_Exp_AP}
	\fl U^{(c)}(z)=\int_{0}^{\infty}(1-e^{\mu t}+ze^{-\mu t})U^{(c)}(1-e^{\mu t}+ze^{-\mu t})\,dF_{\Delta}(t)\\
	+ A\int_{0}^{\infty}\left[\int_{0}^{t}e^{c\mu\sigma^{*}y}(e^{-\mu y}-e^{-\mu t}+ze^{-\mu t})^{c}c\mu\,dy\right]\,dF_{\Delta}(t)-Az^{c}.\nonumber
\end{eqnarray}
Now, let
\begin{equation}\label{eq:Gen_Fun_Devs_AP}
	U^{(c)}_{j}=\frac{1}{j!}\left(\frac{d^{j}U^{(c)}(z)}{dz^{j}}\right)\Big|_{z=1},\qquad(j=0,\dots,c-1).
\end{equation}
Differentiating \eref{eq:Gen_Fun_Exp_AP} $j$ times and evaluating in $z=1$ we have
\begin{equation}\label{eq:Gen_Fun_Diffs0_AP}
	U^{(c)}_{0}=1-\frac{A}{1-\sigma^{*}},
\end{equation}
and
\begin{equation}\label{eq:Gen_Fun_Diffs_AP}
	U^{(c)}_{j}=\frac{\widetilde{\mathcal{F}}(j\mu)}{1-\widetilde{\mathcal{F}}(j\mu)}U^{(c)}_{j-1}-\frac{A}{1-\widetilde{\mathcal{F}}(j\mu)}{c\choose j}\frac{c(1-\widetilde{\mathcal{F}}(i\mu))-j}{c(1-\sigma^*)-j}.
\end{equation}
To solve this linear difference equation, define $C_{0}=1$ and
$$C_j = \prod_{i=1}^j\frac{\widetilde{\mathcal{F}}(i\mu)}{1-\widetilde{\mathcal{F}}(i\mu)},$$
and divide both sides by $C_{j}$. This leads to the result
\begin{equation}\label{eq:Gen_Fun_Diffs2_AP}
	\frac{U^{(c)}_{j}}{C_{j}}=\frac{U^{(c)}_{j-1}}{C_{j-1}}-\frac{A}{C_{j}(1-\widetilde{\mathcal{F}}(j\mu))}{c\choose j}\frac{c(1-\widetilde{\mathcal{F}}(i\mu))-j}{c(1-\sigma^*)-j}.
\end{equation}
Adding these equations for $j=i+1,\dots,c-1$ gives
\begin{equation}\label{eq:Ui_AP}
	\fl U^{(c)}_i = AC_i\sum_{j=i+1}^c\frac{1}{C_i(1-\widetilde{\mathcal{F}}(i\mu))}{c\choose j}\frac{c(1-\widetilde{\mathcal{F}}(i\mu))-j}{c(1-\sigma^*)-j},\quad(r=0,1,\dots,i-1).
\end{equation}

By fixing $i=0$ in \eref{eq:Ui_AP} and using the known value for $U^{(c)}_0$ in \eref{eq:Gen_Fun_Diffs0_AP}, we obtain an explicit formula for the normalising constant $A$ of the form
\begin{equation}\label{eq:A_AP}
	A = \left[\frac{1}{1-\sigma^*} + \sum_{j=1}^c\frac{1}{C_j(1-\widetilde{\mathcal{F}}(i\mu))}{c\choose j}\frac{c(1-\widetilde{\mathcal{F}}(i\mu))-j}{c(1-\sigma^*)-j}\right]^{-1}.
\end{equation}
Finally, to compute the remaining probabilities $\Pi^{(c)}_{0},\dots\Pi^{(c)}_{c-2}$, observe that the generating function can be completely determined using \eref{eq:Gen_Fun_Diffs2_AP} and \eref{eq:A_AP}
\begin{equation}\label{eq:Uz_AP}
	U^{(c)}(z)=\sum_{j=0}^{c-1}U^{(c)}_{j}\cdot(z-1)^{j}.
\end{equation}
Therefore, the remaining probabilities $\Pi^{(c)}_{0},\dots\Pi^{(c)}_{c-2}$ can be expressed as
\begin{equation}\label{eq:Probs_Fin_AP}
	\fl\Pi^{(c)}_{j} = \frac{1}{j!}\left(\frac{d^{j}U^{(c)}(z)}{dz^{j}}\right)\Big|_{z=0}=\sum_{i=j}^{c-1}(-1)^{i-j}{i\choose j}U^{(c)}_{i},\quad(i=0,\dots,c-2).
\end{equation}
\subsection{Distribution of the waiting time}\label{ap:WaitingTime}
Denote by $W_{n}$ the waiting time in the system before service of the $n$th customer and let $F_{W_{n}}(t)=\mathbb{P}[W_{n}\leq t]$. Following the construction of the embedded Markov chain, we know that the $n$th customer finds $Q^{(c)}_{\tau_{n}}$ customers in the system ahead of them. Therefore, if $Q^{(c)}_{\tau_{n}}=j<c$ then the service starts without waiting and if $Q^{(c)}_{\tau_{n}}=j\geq c$ they must wait for $j+1-c$ successive departures, which follow a Poisson process of intensity $c\mu$. From these two facts, it follows that
\begin{equation}\label{eq:FWn_AP}
	\fl F_{W_{n}}(t)=\sum_{j=0}^{c-1}\mathbb{P}[Q^{(c)}_{\tau_{n}}=j]+\sum_{j=c}^{\infty}\mathbb{P}[Q^{(c)}_{\tau_{n}}=j]\int_{0}^{t}e^{-c\mu x}\frac{(c\mu x)^{j-c}}{(j-c)!}c\mu\,dx.
\end{equation}

Now, assuming that $\rho=\lambda/(c\mu)<1$, ensuring that the system converges to a steady-state (as shown in \ref{subap:A3}), and taking limit when $n$ goes to infinity gives \cite{takacs_introduction_1962}
\begin{equation}\label{eq:FWn_Lim_AP}
	\fl F_{W}(t)=\lim_{n\to\infty}F_{W_{n}}(t)=\sum_{j=0}^{c-1}\Pi^{(c)}_{j}+\sum_{j=c}^{\infty}\Pi^{(c)}_{j}\int_{0}^{t}e^{-c\mu x}\frac{(c\mu x)^{j-c}}{(j-c)!}c\mu\,dx.
\end{equation}
Using the fact that $\Pi^{(c)}_{j}=A(\sigma^{*})^{j-c}$ if $j\geq c-1$ we have that
\begin{equation}\label{eq:FWn_AP_1}
\sum_{j=0}^{c-1}\Pi^{(c)}_{j}=1-\sum_{j=c}^{\infty}A(\sigma^{*})^{j-c} = 1-\frac{A}{1-\sigma^{*}},
\end{equation}
and
\begin{eqnarray}\label{eq:FWn_AP_2}
	\fl\sum_{j=c}^{\infty}\Pi^{(c)}_{j}\int_{0}^{t}e^{-c\mu x}\frac{(c\mu x)^{j-c}}{(j-c)!}c\mu\,dx &= \int_{0}^{t}c\mu Ae^{-c\mu x}\sum_{j=c}^{\infty}\frac{(\sigma^{*}c\mu x)^{j-c}}{(j-c)!}\,dx\nonumber\\
	&= c\mu A\int_{0}^{t}e^{-c\mu x(1-\sigma^{*})}\,dx\\
	&= \frac{A}{1-\sigma^{*}}-\frac{A}{1-\sigma^{*}}e^{-c\mu(1-\sigma^{*})t}\nonumber.
\end{eqnarray}
Finally, substituting \eref{eq:FWn_AP_1} and \eref{eq:FWn_AP_2} into \eref{eq:FWn_Lim_AP}, we obtain
\begin{equation}\label{eq:FW_AP}
F_{W}(t)=1-\frac{A}{1-\sigma^{*}}e^{-(1-\sigma^{*})c\mu t}.
\end{equation}
This, in turn, implies that the time an arbitrary customer waits for their service to begin once the system reaches a steady state, denoted by $W$, is such that
\begin{equation}\label{eq:W_AP}
	\mathbb{P}[W>t]=\frac{A}{1-\sigma^{*}}e^{-(1-\sigma^{*})c\mu t}.
\end{equation}
\section{Difference between arriving customer distribution and outside observer distribution}\label{ap:ArrVsOut}
\begin{figure}[h!]
	\begin{center}
		\includegraphics[width = \textwidth]{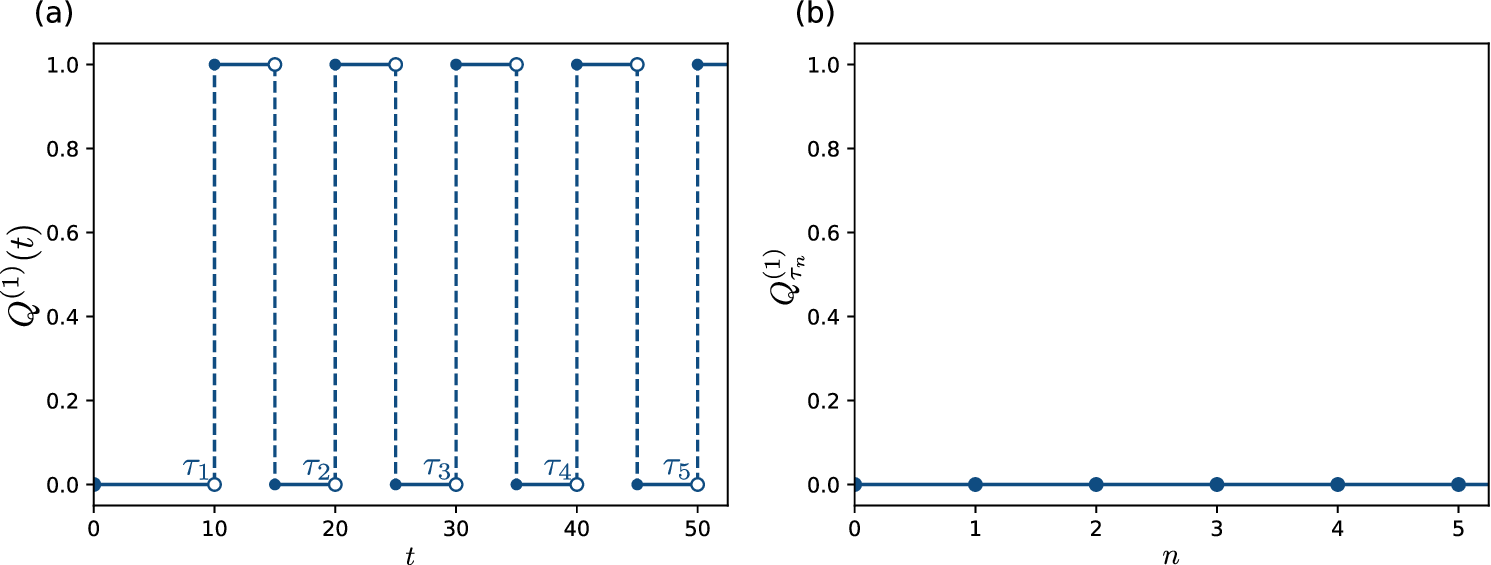}
		\caption{Sample path of the queue length in a $D/D/1$ system where customers arrive every 10 minutes and are served within 5 minutes and its respective embedded Markov chain. (a) Queue length process for the $D/D/1$ system. Each arrival and each departure generate a discontinuity of size one, making the sample path right continuous. The solid dots represent the actual value of the queue length at the time of arrival and departure and the empty dots represent the value just prior to said events. (b) Embedded Markov chain of the queue length process for the $D/D/1$ system.}
		\label{fig:DD1}
	\end{center}
\end{figure}

Consider a queueing system with a deterministic arrival process, a deterministic service time and a single server. We denote this system by $D/D/1$. Assume that customers arrive at the system every 10 minutes and that it takes 5 minutes to fulfil the service of each one of the customers. That is, $\mathbb{P}[\tau_{n}=10n] = 1$, $\mathbb{P}[\Delta_{n}=10] = 1$, and $\mathbb{P}[T_{n}=5] = 1$ for each $n\geq1$; where $\tau_{n}$ is the arrival time of the $n$th customer, $T_{n}$ is the service time of the $n$th customer and $\Delta_{n}$ is the inter-arrival time between customers $n$ and $n-1$. This implies that the arrival rate is $\lambda = 10^{-1}$ and the service rate is $\mu=5^{-1}$, see \Fref{fig:DD1} (a).

Define the embedded Markov chain $\{Q^{(1)}_{\tau_{n}}\}_{n\geq1}$ in the usual way, where $Q^{(1)}_{\tau_{n}}=Q^{(1)}(\tau_{n}^-)$ is the number of customers waiting in the line ahead of the $n$th customer, just prior to the time of their arrival. As customers arrive at the system every 10 minutes and their services only take 5 minutes, each customer fulfils their service before the arrival of the next one, and each arriving customer sees zero customers waiting in the line ahead of them, see \Fref{fig:DD1} (a). Therefore, $Q^{(1)}_{\tau_{n}}=0$ for all $n\geq1$, see \Fref{fig:DD1} (b). It is now clear that the arriving customer distribution is given by $\Pi^{(1)}_{0}=1$ and $\Pi^{(1)}_{n}=0$ for $n\geq1$, since the probability that an arriving customer finds zero customers in the system is one.

On the other hand, in continuous time the system has on average one customer half of the time and zero customers the other half. This implies that the probability that an outside observer at an arbitrary point of time finds one customer in the system is $1/2$ and the probability that they find zero customers in the system is $1/2$. Hence, the outside observer distribution is given by $P^{(1)}_{0}=P^{(1)}_{1}=0.5$ and $P^{(1)}_{n}=0$ for $n\geq2$.

Finally, it is easy to confirm that the arriving customer distribution and the outside observer distribution satisfy the relation stated in \eref{eq:Outside_GMC}:
$$ P^{(1)}_{1}=\frac{1}{2}=\frac{\lambda}{\mu}\Pi^{(1)}_{0}\qquad\mbox{and}\qquad P^{(1)}_{n}=0=\frac{\lambda}{\mu}\Pi^{(1)}_{n-1},\quad\mbox{ for }n\geq2,$$
and
$$P^{(1)}_{0} = 1-P^{(1)}_{1}-\sum_{n=2}^{\infty}P^{(1)}_{n} = 1 - \frac{1}{2} = \frac{1}{2}.$$
\section{Equilibrium statistics for the $G/M/1$}\label{ap:Statistics}
Let $Q^{(1)}_\infty=\lim_{t\to\infty}Q^{(1)}(t)$ be a random variable representing the queue length at steady state, and let $Q^{(1)}_{\tau_{\infty}}=\lim_{n\to\infty}Q^{(1)}(\tau_{n})$ be a random variable representing the queue length at the time of arrival of an arbitrary customer at steady state. That is, $Q^{(1)}_\infty$ has the outside observer distribution of the $G/M/1$ and $Q^{(1)}_{\tau_{\infty}}$ has the arriving customer distribution of the $G/M/1$, $Q^{(1)}_{\tau_{\infty}}\sim$\,Geom$(1-\sigma^*)$. Here, we compute the first and second moments of $Q^{(1)}_\infty$, using the moments of the geometric distribution. We begin by computing the first moment of $Q^{(1)}_\infty$:
\begin{eqnarray*}
	\mathbb{E}[Q^{(1)}_\infty] & = & \sum_{j=0}^\infty jP^{(1)}_j=
	 \frac{\lambda}{\mu}\sum_{j=1}^\infty j(1-\sigma^*)(\sigma^*)^{j-1}\\
	& = & \frac{\lambda}{\mu}\sum_{i=0}^\infty (i+1)(1-\sigma^*)(\sigma^*)^{i}\\
	& = & \frac{\lambda}{\mu}\left[\sum_{i=0}^\infty i(1-\sigma^*)(\sigma^*)^{i}+\sum_{i=0}^\infty(1-\sigma^*)(\sigma^*)^{i}\right]\\
	& = & \frac{\lambda}{\mu}(\mathbb{E}[Q^{(1)}_{\tau_{\infty}}]+1)= \frac{\lambda}{\mu}\left[\frac{\sigma^*}{1-\sigma^*}+1\right]\\
	& = & \frac{\lambda}{\mu(1-\sigma^*)}.
\end{eqnarray*}
Similarly, the second moment of $Q^{(1)}_\infty$ is
\begin{eqnarray*}
	\mathbb{E}[(Q^{(1)}_\infty)^2] & = & \sum_{j=0}^\infty j^2P^{(1)}_j
	= \frac{\lambda}{\mu}\sum_{j=1}^\infty j^2(1-\sigma^*)(\sigma^*)^{j-1}\\
	& = & \frac{\lambda}{\mu}\sum_{i=0}^\infty (i+1)^2(1-\sigma^*)(\sigma^*)^{i}\\
	& = & \frac{\lambda}{\mu}\left[\sum_{i=0}^\infty i^2(1-\sigma^*)(\sigma^*)^{i}+2\sum_{i=0}^\infty i(1-\sigma^*)(\sigma^*)^{i}+\sum_{i=0}^\infty(1-\sigma^*)(\sigma^*)^{i}\right]\\
	& = & \frac{\lambda}{\mu}(\mathbb{E}[(Q^{(1)}_{\tau_{\infty}})^2]+2\mathbb{E}[Q^{(1)}_{\tau_{\infty}}]+1)\\
	& = & \frac{\lambda}{\mu}\left[\frac{\sigma^*+(\sigma^*)^2}{1-\sigma^*}+\frac{2\sigma^*}{1-\sigma^*}+1\right]\\
	& = & \frac{\lambda}{\mu}\left[\frac{1+\sigma^{*}}{(1-\sigma^*)^{2}}\right].
\end{eqnarray*}

\section{Survival function of the superposition of inter-arrival times}\label{ap:Renewal}
Let $\{\Delta_{n}\}_{n\ge0}$ be a sequence of nonnegative random variables with distribution $F_{\Delta}(x)$. We define the renewal process as the sequence $\{S_{n}\}_{n\geq0}$ defined by $S_{0}=0$ and $S_{n}:=\Delta_{1}+\dots+\Delta_{n}$. Generally, the random variables $\Delta_{n}$ are known as the \textit{inter-arrival times}, and the random variables $S_{n}$ are known as the \textit{renewal epochs}. Thus, we can define the counting process $\{N(t):t\geq0\}$ associated with the renewal process $\{S_{n}\}_{n\geq0}$ as
\begin{equation}\label{eq:CountingProcess}
	N(t) := \sup\{n\geq0:S_{n}\leq t\}.
\end{equation}
It follows that $S_{N(t)}\leq t$ and $S_{N(t)+1}\geq t$ almost surely. Therefore, we can define the following lifetime processes associated with the renewal process. We define the \textit{forward recurrence time} as $A(t):=t-S_{N(t)}$ for $t\geq0$. This represents the time elapsed from the last renewal epoch up to time $t$. We also define the \textit{backward recurrence time} as $B(t):=S_{N(t)+1}-t$ for $t\geq0$. This represents the time elapsed between observation $t$ and the next renewal epoch. With these two definitions we can define the \textit{recurrence time} as $C(t):=A(t)+B(t)=\Delta_{N(t)+1}$ for $t\geq0$. This represents the time between two renewal epochs. We now state the key result of the limiting distributions of the lifetime processes.
\begin{theorem*}[Theorem 1.18 from \cite{mitov_renewal_2014}]
Suppose that the mean inter-arrival time, denoted by $\gamma$, is finite $(0<\gamma<\infty)$ and $F_{\Delta}(\cdot)$ is a non-lattice distribution.
\begin{enumerate}
\item For $x\geq0$
\begin{equation}\label{eq:LimitAt}
	\lim_{t\to\infty}\mathbb{P}(A(t)\leq x) = \frac{1}{\gamma}\int_{0}^{x}(1-F_{\Delta}(u))du.
\end{equation}
\item For $x\geq0$
\begin{equation}\label{eq:LimitBt}
	\lim_{t\to\infty}\mathbb{P}(B(t)\leq x) = \frac{1}{\gamma}\int_{0}^{x}(1-F_{\Delta}(u))du.
\end{equation}
\item For $x\geq0$
\begin{equation}\label{eq:LimitVt}
	\lim_{t\to\infty}\mathbb{P}(C(t)\leq x) = \frac{1}{\gamma}\int_{0}^{x}u\,dF_{\Delta}(u).
\end{equation}
\end{enumerate}
\end{theorem*}

We denote the limiting backward recurrence time distribution defined in \eref{eq:LimitBt} as $F_{B}(x)$ and the limiting backward recurrence time survival function as $Q_{B}(x) = 1 - F_{B}(x)$. It can be proven that
\begin{equation}\label{eq:LimitBtResult1}
	Q_{B}(x) = \frac{1}{\gamma}\int_{x}^{\infty}Q_{\Delta}(u)du,
\end{equation}
where $Q_{\Delta}(x)=1-F_{\Delta}(x)$ is the survival function of the inter-arrival times. Additionally, denoting the limiting backward recurrence time density, it is easy to corroborate that
\begin{equation}\label{eq:LimitBtResult2}
	f_{B}(x)=-\frac{d}{dx}Q_{B}(x) = \frac{Q_{\Delta}(x)}{\gamma}.
\end{equation}

Having all the above results for a single renewal process, we now assume that we have a family of inter-arrival time sequences
\begin{equation}\label{eq:IATsSuperposition}
	\mathcal{I} := \bigcup_{i\geq1}\{\Delta_{n}^{(i)}\}_{n\ge0},
\end{equation}
where $F^{(i)}_{\Delta}(t)$ is the distribution of the $i$th inter-arrival time sequence. Each inter-arrival time sequence independently defines a renewal process $\{S_{n}^{(i)}\}_{n\geq0}$ and its associated counting process $\{N(t)^{(i)}:t\geq0\}$. We consider $\{\mathcal{S}_{n}^{(M)}\}_{n\geq0}$ as the sequence of renewal epochs obtained by considering the union of the first $M$ renewal processes $\{S_{n}^{(i)}\}_{n\geq0}$ and arranging them in increasing order. This represents the renewal epochs, or customer arrivals at a target, coming from any of $M$ different sources or searchers. We can define a counting process associated with $\{S_{n}^{(i)}\}_{n\geq0}$ in the following simple manner
\begin{equation}\label{eq:CountingProcessSuperposition}
	\mathcal{N}_{M}(t):= \sum_{i=1}^{M}N(t)^{(i)},\qquad t\geq0.
\end{equation}

Assuming that all of the $M$ renewal processes have the same inter-arrival time distribution, $F_{\Delta}$, with finite mean inter-arrival time $(0<\gamma<\infty)$, we find that the mean inter-arrival time for $\{\mathcal{S}_{n}^{(M)}\}_{n\geq0}$ is simply $\gamma M^{-1}$. Now, the lifetime processes associated with the combined sequence of renewal epochs $\{\mathcal{S}_{n}^{(M)}\}_{n\geq0}$ take the form
\begin{equation}\label{eq:AtSuperposition}
	\mathcal{A}_{M}(t):= \min\{A^{(1)}(t),\dots,A^{(M)}(t)\},\qquad t\geq0,
\end{equation}
and
\begin{equation}\label{eq:BtSuperposition}
	\mathcal{B}_{M}(t):= \min\{B^{(1)}(t),\dots,B^{(M)}(t)\},\qquad t\geq0.
\end{equation}

We can use these definitions and the results for the limiting distributions of the lifetime processes for a single renewal process to determine the inter-arrival time survival function for the superposition of $M$ identical, independent renewal processes. First, we consider the limiting backward recurrence times $\mathcal{B}_{M} = \lim_{t\to\infty}\mathcal{B}_{M}(t)$, where we consider the limit $B^{(i)} = \lim_{t\to\infty}B^{(i)}(t)$ for each $i\in\{1,\dots,M\}$. Using the definition of the backward recurrence time of the superposition process \eref{eq:BtSuperposition} and the result of the limiting distribution for the backward recurrence time \eref{eq:LimitBt}, we obtain
\begin{equation}\label{eq:BtLimitSuperposition}
	\fl\mathbb{P}(\mathcal{B}_{M}>x) = \mathbb{P}\left(\min_{1\leq i\leq M}B^{(i)}>x\right) = \prod_{i=1}^{M}\mathbb{P}(B^{(i)}>x)=\left(\frac{1}{\gamma}\int_{x}^{\infty}Q_{\Delta}(u)du\right)^{M}.
\end{equation}
Denoting the limiting backward recurrence time density for the superposition process by $f_{\mathcal{B}_{M}}$ we have
\begin{equation}\label{eq:BtSuperpositionDensity}
	f_{\mathcal{B}_{M}}(x) = -\frac{d}{dx}\mathbb{P}(\mathcal{B}_{M}>x) = Q_{\Delta}(x)\frac{M}{\gamma}\left(\frac{1}{\gamma}\int_{x}^{\infty}Q_{\Delta}(u)du\right)^{M-1}.
\end{equation}
Now, using the fact that the limiting backward recurrence time density equals the survival function of the inter-arrival times over the mean inter-arrival time \eref{eq:LimitBtResult2} and denoting the survival function of the inter-arrival times for the superposition process by $Q_{M}$ we finally have that
\begin{equation}\label{eq:SurvivalSuperposition}
	Q_{M}(x) = Q_{\Delta}(x)\left(\frac{1}{\gamma}\int_{x}^{\infty}Q_{\Delta}(u)du\right)^{M-1}.
\end{equation}

\section*{References}

\providecommand{\newblock}{}

\end{document}